\documentclass[aps,pra,twocolumn,superscriptaddress]{revtex4}
\usepackage[english]{babel}
\usepackage[T1]{fontenc}
\usepackage{amsmath,amsfonts,amssymb,float,graphicx,epsfig,color,bbm,bm,txfonts}
\newcommand{\id}{{\sf 1 \hspace{-0.3ex} \rule{0.1ex}{1.52ex}\rule[-.01ex]{0.3ex}{0.1ex}}}
\newcommand{\ket}[1]{\left | #1 \right\rangle}
\newcommand{\bra}[1]{\left \langle #1 \right |}

\renewcommand{\epsilon}{\varepsilon}
\begin{document}
\title{Experimental Demonstration of non-Markovian Dynamics via a Temporal Bell-like Inequality}
\author{A. M. Souza}
\affiliation{Centro Brasileiro de Pesquisas F\'isicas, Rua Dr. Xavier Sigaud 150, 22290-180 Rio de Janeiro, Rio de Janeiro, Brazil}
\author{J. Li}
\affiliation{Centre for Theoretical Atomic, Molecular and Optical Physics,
School of Mathematics and Physics, Queen's University Belfast, BT7 1NN, United Kingdom}
\author{D. O. Soares-Pinto}
\affiliation{Instituto de F\'isica de S\~ao Carlos, Universidade de S\~ao Paulo, Caixa Postal 369, 13560-970 S\~ao Carlos, S\~ao Paulo, Brazil}
\author{R. S. Sarthour}
\affiliation{Centro Brasileiro de Pesquisas F\'isicas, Rua Dr. Xavier Sigaud 150, 22290-180 Rio de Janeiro, Rio de Janeiro, Brazil}
\author{I. S. Oliveira}
\affiliation{Centro Brasileiro de Pesquisas F\'isicas, Rua Dr. Xavier Sigaud 150, 22290-180 Rio de Janeiro, Rio de Janeiro, Brazil}
\author{S. F. Huelga}
\affiliation{Institut f\"ur Theoretische Physik, Albert-Einstein-Allee 11, Universit\"at Ulm, D-89069 Ulm, Germany}
\author{M. Paternostro}
\affiliation{Centre for Theoretical Atomic, Molecular and Optical Physics, School of Mathematics and Physics, Queen's University Belfast, BT7 1NN, United Kingdom}
\affiliation{Institut f\"ur Theoretische Physik, Albert-Einstein-Allee 11, Universit\"at Ulm, D-89069 Ulm, Germany}
\author{F. L. Semi\~ao}
\affiliation{Centro de Ci\^encias Naturais e Humanas, Universidade Federal do ABC, 09210-170, Santo Andr\'e, S\~ao Paulo, Brazil}
\begin{abstract}
{All physical systems are, to some extent, affected by the environment they are surrounded by. For this reason, a clear understanding of the physical laws governing the evolution of classical and quantum open systems is of fundamental importance to a diverse community in physics. In this context, the ability to determine whether or not the evolution of a given open system may be well described by a classical memoryless (Markovian) model instead of a memory-keeping non-Markovian process (either classical or quantum) is of quite general interest~\cite{petruccione,RivasHuelga}. Remarkably, non-Markovianity of quantum evolutions is emerging as a resource for quantum technological applications~\cite{applications,applications2,applications4,applications5,applications6} and the key to characterise the nature of fundamental (charge and energy) transport processes in biological aggregates~\cite{SusanaNatPhys,Aspuru} and complex nanostructures~\cite{Li}. Special forms of Bell-like inequalities in time serve as useful tools to detect deviations from any possible classical Markovian description~\cite{nori_lambert1,EmaryReview}. 

In this paper, we assess non-Markovianity of a quantum open-system dynamics through the violation of one of such inequalities using  a controllable Nuclear Magnetic Resonance (NMR) system~\cite{NMRbooks}. We establish a clear relation between the violation of the addressed temporal Bell-like inequality and the non-divisibility of the effective evolution of our system, which we fully characterize experimentally in a broad range of experimentally controllable situations.}

\end{abstract}
\maketitle

Temporal Bell-like inequalities such as those originally proposed by Leggett and Garg embody methods to investigate the possibility to witness {\it macroscopic coherence} in the state of a system~\cite{leggett_garg}. Starting from the classically valid assumptions that, in principle, measurements can be made on a system without affecting its subsequent evolution (known as the {``non-invasive measurability''} assumption) and, at any instant of time, the system itself will be in a well-defined state among those it has available (embodying the assumption of {``macroscopic realism''}), the inequalities set in Ref.~\cite{leggett_garg} provide a benchmark for any dynamics conforming to our classical intuition. The violation of such inequalities, which are built by combining time correlators of a suitably chosen observable of the system, rule out the framework defined by the two assumptions above and that is commonly intended as macrorealism, have been recently reported in setups based on linear optics \cite{pnas,Dressel}, NMR~\cite{lg_cbpf, Athalyse}, superconducting quantum circuits \cite{laloy}, spin impurities in silicon~\cite{Knee} and a nitrogen-vacancy defect in diamond~\cite{Waldherr}. 

Going somehow beyond the framework of macro realism, it has been found that time correlators can also be arranged in special inequalities that help detecting deviations from a classical memoryless dynamical map~\cite{petruccione}. One of such inequalities can be cast into the form~\cite{nori_lambert1}  
\begin{equation}
\label{extended}
|L_Q(t)|=|2\langle \hat Q(t)\hat Q(0)\rangle-\langle \hat Q(2t)\hat Q(0)\rangle|\le Q_{max}\langle \hat Q(0)\rangle
\end{equation}
with $\hat Q$ a suitable observable of the system, $\langle \hat Q(t_1)\hat Q(t_2)\rangle$ its two-time autocorrelation function, and $Q_{max}$ the maximum taken by the expectation value $\langle \hat Q\rangle$ (calculated over the state of the system at hand) and that should simply be interpreted as a normalization factor. Eq.~\eqref{extended} holds under the assumption of memoryless (or Markovian) evolution of the system. 
Under proper conditions on the time-dependence of the autocorrelation functions \cite{nori_lambert1}, and retaining the non-invasive nature of the measurements of ${\hat Q}$, this inequality embodies a test of macrorealism and is fully equivalent to the Leggett-Garg one~\cite{leggett_garg,nori_lambert1}. However, in general, the two inequalities are not the same and we shall refer to Eq.~(\ref{extended}) as the {\it extended Leggett-Garg (LG) inequality}. The right-hand side of this inequality is calculated assuming a classical stochastic Kolmogorov  framework~\cite{petruccione}. In a nutshell, this implies that the dynamics of the system can be described by the classical stochastic map $\dot{\bm p}(t)={\cal M}(t){\bm p}(t)$, where ${\bm p}(t)$ is the vector of  single-time probabilities for the stochastic process at hand, and ${\cal M}(t)$ is a time-dependent matrix whose entries satisfy the Kolmogorov conditions ${\cal M}_{ij}\ge0$ (for $i\neq{j}$), ${\cal M}_{ii}\le0$, and $\sum_{j}{\cal M}_{ij}=0$ (for any $i$)~\cite{petruccione}. When the bound imposed on the two-time correlators by Eq.~\eqref{extended} is violated, the system fails to satisfy such assumption and its stochastic dynamics departs from a classical Markovian one~\cite{EmaryReview}.

\begin{figure}[t]
\begin{center}
\includegraphics[width=\columnwidth]{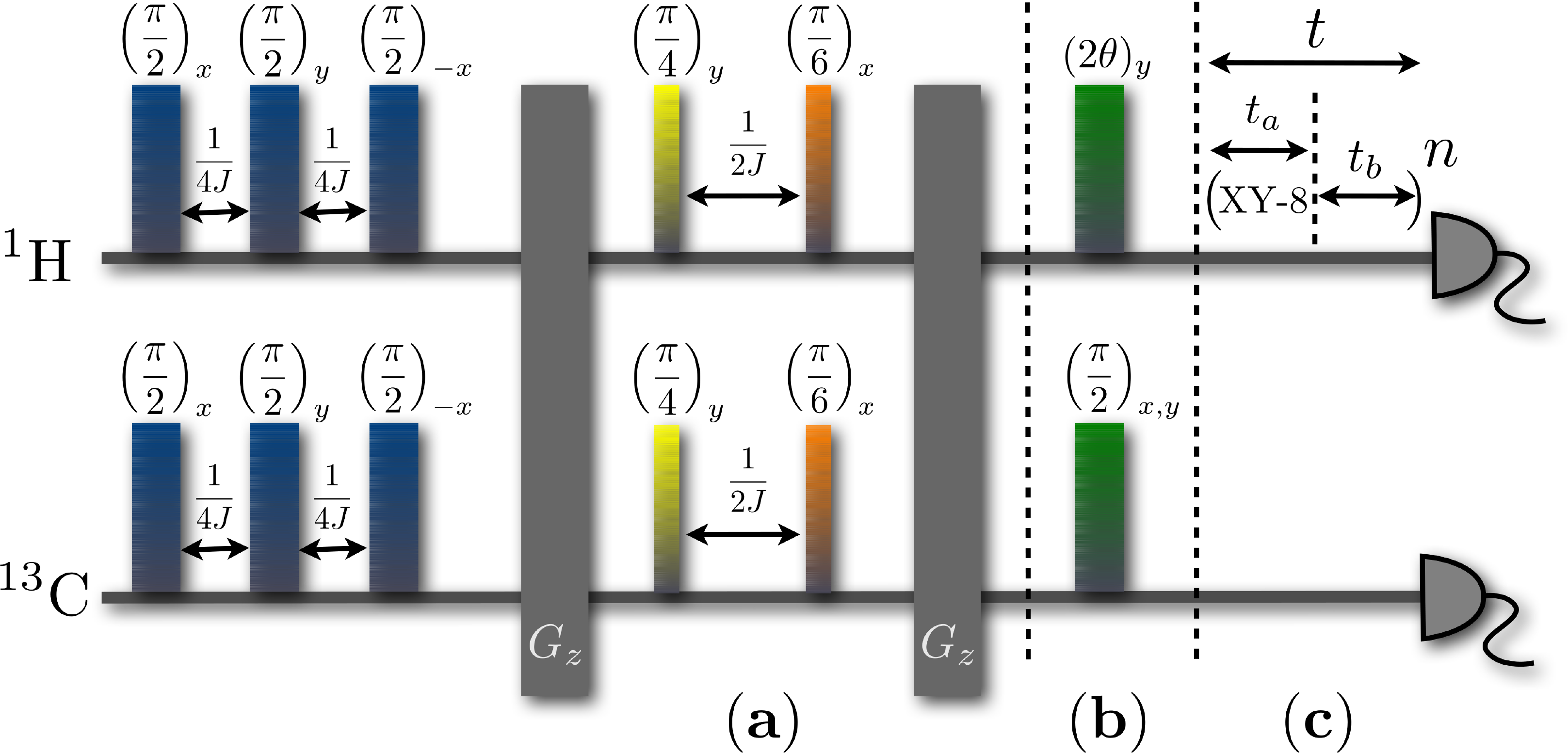}
\end{center}
\caption{{Sketch of the rf pulse-sequence for state preparation and  evolution}. 
Part {\bf (a)} and {\bf (b)} show the steps for the initial state preparation. Starting from a thermal equilibrium state of the $S$ and $E$ qubits, we apply the pulse sequence shown in part {\bf (a)} of the figure to generate the state $\tilde\rho = {(1-\epsilon)}\openone/{4}+ \epsilon\rho_{1}$ with $\rho_{1} = |00\rangle\langle 00|_{SE}$. We use the notation-shortcut $\left(\theta\right)_{\alpha}$ to indicate a qubit rotation by the angle $\theta$ about the direction $\alpha=\pm x,\pm y,\pm z$. Moreover, $\frac{1}{4J}$ and $\frac{1}{2J}$ stand for length of the evolution of the qubits under their mutual coupling and $G_{z}$ for a pulsed gradient field applied over the qubits. Such field dephases the state of $S$ and $E$ leaving only the  diagonal elements of the density matrix unaffected. This results in state $\tilde\rho$. In part {\bf (b)} wet make use of two more pulses to create the desired state of the register. The $^{13}$C nuclear spins are prepared in the state $|+\rangle_{S}$ or $|-\rangle_{S}$, depending on the phase of the rf pulse it is subjected to. On the other hand, the $^{1}$H spins are prepared in state $|\theta\rangle_{E} = \cos\theta |0\rangle_{E} + \sin\theta |1\rangle_{E}$ thanks to a properly chosen single-qubit rotation $(2\theta)_y$. Part {\bf (c)} sketches how to implement the effective coupling between the nuclear spins for a time $t$. Note that this part of evolution is divided in two intervals, each of length $t_a$ and $t_b$. During $t_a = (1 - J_{\rm eff}/J)\,t$ a decoupling sequence is applied to the $^1$H nuclear spin so to refocus the evolution under the Ising-like coupling. As the decoupling sequence must work for every initial state, we choose the sequence XY-$8$ to accomplish this result (cf. Ref.~\cite{amsouzaDD} for details). During $t_b = (J_{\rm eff} / J)\,t$ the system evolves freely. The net effect at $t = t_a + t_b$ is to implement an evolution under the effective coupling strength $J_{\rm eff}$. The $n$-fold repetition of the XY-$8$ scheme maintains the desired effective coupling during the whole time-window of the evolution of the qubits.}
\label{pulsesequence}
\end{figure}

\begin{figure*}[ht]
\begin{center}
\includegraphics[width=2\columnwidth]{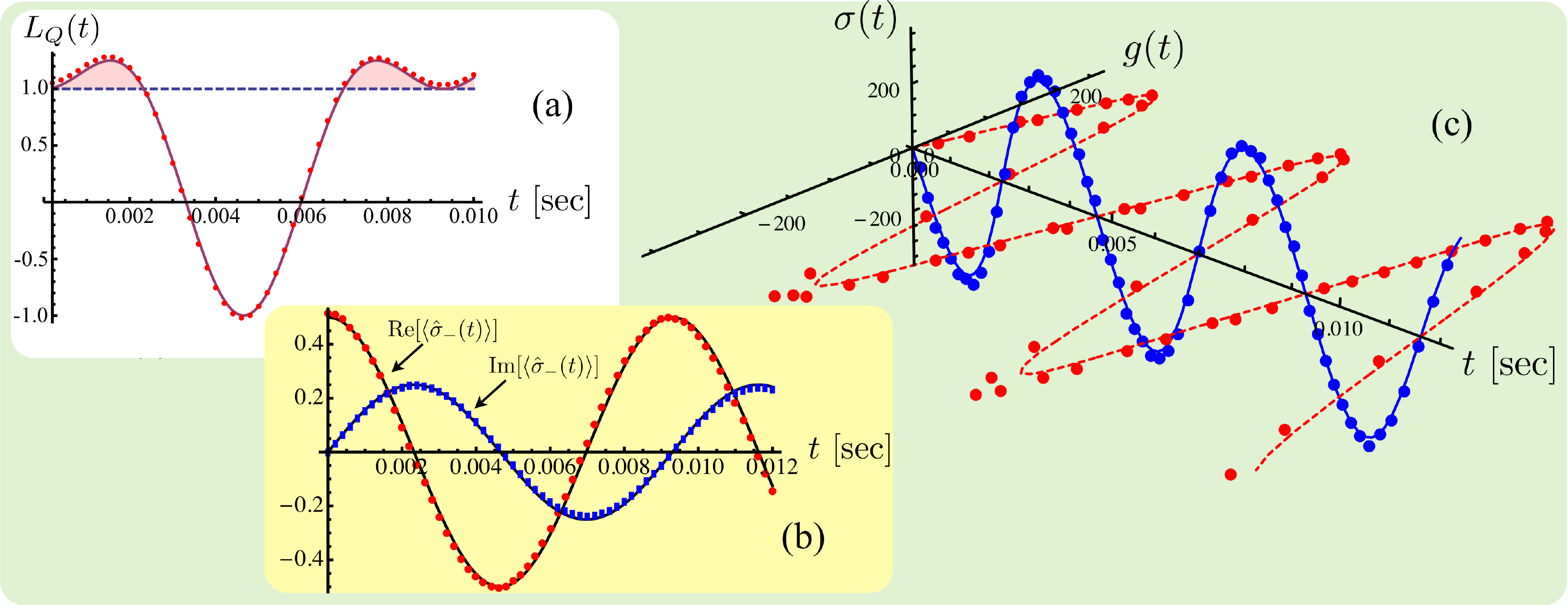}
\end{center}
\caption{{Violation of the extended LG inequality, tomography of the open-system dynamics, and indicators of non-Markovianity}. In all panels the dots identify the experimental data while the solid or dashed lines show the behaviour of the theoretical function. {\bf (a)} Generalized Leggett-Garg inequality. We plot the generalized Leggett-Garg function $L_Q(t)$ [cf. Eq.~\eqref{elg}] against the evolution time $t$ for $\hat Q=|+\rangle_S\langle+|$, $J=215.06$Hz, $\theta=\pi/3$ and the system initially prepared in state $\ket{+}_S$. The shadowed regions identify the violation of the inequality stated in Eq.~(\ref{elg}). {\bf (b)} We show the time behavior of the real (red circles) and imaginary part (blue squares) of the transverse magnetization for a system initially prepared in the $\ket{+}_S$ state. We used the same parameters as in panel {\bf (a)}. {\bf (c)} We plot the time-derivative of the trace distance $\sigma(t)$ (blue circles) and the temporal behavior of the effective time-dependent dephasing rate $g(t)$ (red ones) for the same parameters used in panel {\bf (a)}. Negative values of $g(t)$ are perfectly correlated with positive values for $\sigma(t)$, as detailed in the main text.}
\label{centralone}
\end{figure*}

In order to test the inequality experimentally and in a controlled setting, we have used the NMR set-up consisting of $^1$H and $^{13}$C nuclear spins in a liquid sample of Carbon-enriched chloroform in deuterated solvent~\cite{cbpfbook}. In what follows, we will identify the system of interest $S$ with the $^{13}$C nuclear spin and the environmental two-level system $E$ with the $^1$H ones. The nuclear spins of the two atomic species have Larmor frequency $\omega_E/2\pi \approx 500$ MHz and $\omega_S/2\pi \approx 125$ MHz, and each can be seen as a two-level system~\cite{cbpfbook}. Mutual interaction between them is  enforced by an Ising-like term of the form $\hat H_{I} = 2\pi J\hat I^S_z\hat I^E_z$ with $J$ the corresponding coupling constant. The overall Hamiltonian, written in the laboratory frame, thus reads~\cite{NMRbooks}
\begin{equation}\label{H}
\hat H=-\omega_S\hat I_z^S-\omega_E\hat I_z^E+\hat H_I
\end{equation}
with $\hat I^{S,E}_z = \hbar \hat\sigma^{S,E}_z/2$ and $\hat{\sigma}^{S,E}_z$ the Pauli $z$ operator. 

In the reminder of this paper, we will work in the interaction picture with respect to the free Hamiltonian of both $S$ and $E$, so that only the Ising-like term will be retained. At room temperature, the ratio $\epsilon_{S,E} = \hbar\omega_{S,E}/4k_{B}T$ between the energy gap of each two-level system and the thermal energy is typically of the order of $10^{-5}$ (here $k_B$ is the Boltzmann constant and $T$ the operating temperature). This means that the density matrix of the system can be written in the high-temperature approximation as $\rho = \openone/4 + \epsilon_S\Delta\rho$. The term $\Delta\rho$ is known as the deviation matrix and contains all the information about the state of the sample~\cite{cbpfbook}. A combination of a series of radio-frequency(rf) pulses of appropriate length, phase and amplitude, all of which embody unitary transformations on $\Delta\rho$, joint evolutions under the spin-spin interaction given above, and temporal/spatial averaging procedures~\cite{VanderChuang}, allow us to manipulate the state of the sample with an excellent degree of control. The experimental reconstruction of such state requires the application of a specific sets of rf pulses to $\Delta\rho$~\cite{cbpfbook}. Fig.~\ref{pulsesequence} shows the rf pulse-sequence used in this work to prepare the initial state of the sample and evolve it. Part {\bf (a)} and {\bf (b)} of the sequence are used to prepare the initial states $|+\rangle$ and $|-\rangle$ of $S$, i.e., the eigenstates of the Pauli $x$ operator $\hat\sigma^S_x$ associated with eigenvalue $\pm1$ (the first is obtained when the phase of the rf pulse shown in green is $y$ and the second when it is $x$), and $|\theta\rangle = \cos\theta |0\rangle + \sin\theta |1\rangle$ for the environment. 
After the preparation of the respective states, $S$ and $E$ are made to evolve under the influence of ${\hat H}$ with a specific value of the coupling strength and for a time interval $t = n\,\tau$ (for $n = 0, 1, 2, 3, \dots$ and $\tau=250\mu$s). The actual value of $J$ (determined by the nature of the sample) can be decreased to virtually any value $J_{\rm eff}<J$ by using a decoupling pulse sequence, named XY-8, to refocus  the Ising-like coupling and thus yield the desired value of the interaction strength~\cite{amsouzaDD}.

In order to show that the dynamics experienced by the system only (i.e., the one obtained by tracing out $E$) indeed violates the generalised LG inequality, thus deviating from the assumptions of a classical Markovian dynamics,  we have prepared $S$ in $\ket{+}_S$ and let it evolve together with the environmental qubit according to Eq.~\eqref{H}. Moreover, we have chosen to consider the system's observable $\hat Q=|+\rangle\langle +|_S$, so that $Q_{max}=\langle Q(0)\rangle=1$ and Eq.~eqref{extended} can be written in the simplified form
\begin{equation}
\label{elg}
|L_Q(t)|=|2\langle \hat Q(t)\hat Q(0)\rangle-\langle \hat Q(2t)\hat Q(0)\rangle|\le 1.
\end{equation}
The two-time correlation functions of our choice for $\hat Q$ can be straightforwardly evaluated from the evolved density matrices of $S$. We have thus reconstructed the evolved $S$-$E$ density matrices at various instants of their evolution using standard experimental techniques for quantum state tomography in NMR~\cite{cbpfbook}. From these data, the reduced state of $S$, and thus the two-time correlation function, have been easily extracted. As shown in the Appendix, the average fidelity between the theoretical states and the reconstructed  density matrices is as large as $0.992\pm0.003$ across the whole sample of states probed in our experiment. Fig.~\ref{centralone} {\bf (a)} shows the experimental behavior of the extended LG function within an evolution time of 10ms for $J=215.06$Hz and the environment initially prepared in the state $|\theta=\pi/3\rangle_E$. Besides showing a remarkable agreement with the theoretical predictions (see the Appendix), the data shown in Fig.~\ref{centralone} {\bf (a)} reveal the considerable unsuitability of a classical memoryless picture for the description of the dynamics of $S$, therefore falsifying experimentally any classical Markovian picture to describe its evolution. To the best of our knowledge, this is the first experimental test of Eq.~\eqref{elg} in a controlled scenario.

Two important features should be noticed at this point: First, as shown in the Appendix and verified in our experiment, the violation of Eq.~\eqref{elg} in our model is independent of the environmental state as the two-time correlation function of $\hat{Q}$ does not bear any dependence on $\theta$. Second, the falsification of the extended LG inequality Eq.~\eqref{extended} [or Eq.~\eqref{elg}] does not provide a conclusive statement on the reasons behind the deviation of our experimentally inferred $L_Q(t)$ function from a  classical Markovian picture. {Classical non-Markovian} as well as a {quantum Markovian} maps may equally violate an extended LG inequality, and we now aim at shedding light on the reasons behind the behavior revealed by our experimental observations. In order to do so, we start proving that the dynamics undergone by $S$ is strongly non-Markovian according to some quantitative measure. 

We remark that the temporal evolution of $S$ can be described by using the theory of quantum channels in terms of the dephasing map~\cite{nielsenchuang} 
\begin{equation}
\label{channel}
\rho_S(t)=\text{Tr}_E[\hat{\cal U}(t)\rho_{SE}(0)\hat{\cal U}^\dag(t)]=
\begin{pmatrix}
\rho_{00} & \eta_\theta(t)\rho_{01} \\
 \eta_\theta^*(t)\rho_{10} & \rho_{11}
\end{pmatrix},
\end{equation}
where we have introduced the time propagator of the $S$-$E$ system ${\cal U}(t)=\exp[-i\hat H t]$, the elements of the system's density matrix $\rho_{ij}={}_S\!{\langle{i}|\rho_S(0)\ket{j}_S}~(i,j=0,1)$, and the $E$-dependent dephasing rate $\eta_\theta(t)=e^{-i\pi Jt} \cos ^2(\theta )+e^{i \pi  J t}\sin^2(\theta)$. Eq.~\eqref{channel} is well suited for making a comparison with the results of our experiment. By measuring the transverse magnetization $\langle\hat\sigma^S_-(t)\rangle$ of the system qubit at various instants of the evolution, we have determined the trend followed experimentally by $\eta_\theta(t)$ for given choices of the environmental state and the Ising coupling rate. The data have been found to be in excellent agreement with the theoretical predictions, as shown in Fig.~\ref{centralone} {\bf (b)} for $S$ prepared in $\ket{+}_S$ and $\theta=\pi/3$ (two more choices of initial system's state are discussed in the Appendix).

Notwithstanding the closed form of the evolved state of the system, we would like to write $\rho_S(t)$ in a form that can be readily connected to the non-Markovian features of the dynamics it is undergoing. To this end, we exploit the formal apparatus discussed in Refs.~\cite{anderson,smirne} for deriving a time-local master equation starting from Eq.~\eqref{channel} and find (see the Appendix)
\begin{equation}
\label{ME}
\partial_t\rho_{S}(t)=if(t)[\hat\sigma_z^S,\rho_S(t)]+(g(t)+\gamma)[\hat\sigma_z^S\rho_S(t)\hat\sigma^S_z-\rho_S(t)]
\end{equation}
with the effective Bohr frequency and dephasing rate
\begin{equation}
\label{fandg}
\begin{aligned}
f(t)&=\frac{2 \pi  J \cos (2 \theta )}{3+2 \cos (4 \theta ) \sin ^2(\pi  J t)+\cos (2 \pi  J t)},\\
g(t)&=\frac{\pi  J \sin ^2(2 \theta ) \sin (2 \pi  J t)}{3+2 \cos (4 \theta ) \sin ^2(\pi  J t)+\cos (2 \pi  J t)},
\end{aligned}
\end{equation}
while $\gamma$ embodies the standard dephasing rate of the system's state due to the fluctuating environment surrounding the NMR sample. 
An experimental estimate leads to the value $\gamma\simeq{5}$Hz. As we have so far operated at $J\gg{\gamma}$ and within an evolution window much shorter than $\gamma^{-1}$, we could safely neglect any influence of such environmental mechanism, as also demonstrated by the excellent agreement between theoretical predictions and actual experimental data shown in Fig.~\ref{centralone}. 
Eq.~\eqref{ME} shows very clearly that the system-environment coupling gives rise to a time-dependent dephasing mechanism that affects the coherence of the $S$ state as $\rho_{01}\to\rho_{01}e^{-2\gamma t+2\int^t_0\phi(t')dt'}$ with $\phi(t)=if(t)-g(t)$, leaving the populations unaffected. Moreover, the negativity of the total time-dependent dephasing rate $\gamma+g(t)$ at some instant of time during the evolution of $S$ would signal the break-down of the divisibility condition of the underlying dynamical map and thus its non-Markovian nature, as stated by the criterion proposed in Ref.~\cite{Rivas}. 

Such formulation of the system's dynamics is useful in two respects: first, it allows us to link the behavior of the transverse magnetization to the effective dephasing rate of the system's evolution. The comparison between the experimentally measured transverse magnetization and $\phi(t)$ enables the identification of the trend followed by $f(t)$ and $g(t)$. Second, it provides an experimental route to the characterization of the non-Markovianity of Eq.~\eqref{ME}. Indeed, using the data for the transverse magnetization, we have inferred the full experimental form of $\phi(t)$ using the relations $g(t)=(1/2)\text{Re}[\langle\dot{\hat\sigma}_-(t)\rangle/\langle\hat{\sigma}_-(t)\rangle]$ and $f(t)=(1/2)\text{Im}[\langle\dot{\hat\sigma}_-(t)\rangle/\langle\hat{\sigma}_-(t)\rangle]$. 

Using this approach, we have performed an effective tomography of the master equation followed by the system qubit $S$. In Fig.~\ref{centralone} {\bf (c)} [dashed line] we report the comparison between the experimentally inferred $g(t)$ and the second line of Eq.~\eqref{fandg}. The data demonstrate unambiguously the occurrence of ample regions of negativity of $g(t)$ for the value of the relevant parameters entering Eq.~\eqref{ME}, and thus the non-Markovian character of the corresponding dynamics of $S$. Such conclusions are corroborated by an analysis based on the measure for non-Markovianity proposed in Ref.~\cite{breuer}, where positivity of the quantity $\sigma(t)=\partial_t||\rho_{S,1}(t)-\rho_{S,2}(t)||$ for a pair of states $\rho_{S,1(2)}$ of the system witnesses dynamical non-Markovianity. Here, we have introduced the trace-norm $||A||=\text{Tr}[\sqrt{\hat A^\dag A}]$ of a generic matrix $A$~\cite{nielsenchuang}. In our experiment we have taken $\rho_{S,1}(0)=\ket{+}_S\!\bra{+}$ and $\rho_{S,2}(0)=\ket{-}_S\!\bra{-}$, a choice that guarantees the maximisation of $\sigma(t)$ (and thus of the degree of non-Markovianity), which reads
\begin{equation}
\sigma(t)=-g(t){\sqrt{3+2 \cos (4 \theta ) \sin ^2(\pi  J t)+\cos (2 \pi  J t)}}.
\end{equation}
Although the two criteria are in general inequivalent, for pure dephasing mechanisms the conclusions drawn from the assessment of non-divisibility of a map and the evolution of the trace-distance witness are identical~\cite{Haikka}. The assessment of $\sigma(t)$ should thus be taken as an important and interesting consistency check heralding the presence of memory effects. 

In Fig.~\ref{centralone} {\bf (c)} we show that the experimentally inferred dephasing rate $g(t)$ is in full opposition of phase with respect to the $\sigma(t)$ calculated using the states of $S$ reconstructed through experimental quantum state tomography. The agreement between experimental data and theoretical predictions allows us to claim that  the deviation of the system's dynamics from a Markovian picture concurs to the experimental falsification of Eq.~\eqref{elg}. 

In this respect, it is interesting to notice that the reduced dynamics of the system (initialised in state $\ket{+}_S$) is unable to violate the original LG inequality~\cite{leggett_garg} when either $\ket{+}_S\!\bra{+}$ or $\hat\sigma^S_x$ are chosen as observables and the environment is prepared in $\ket{\theta=\pi/3}$. In fact, while $\ket{+}_S\!\bra{+}$ is unsuited for the violation of the original LG inequality regardless of the state of $E$, the second choice of observable would falsify any macro realistic picture (should the measurements of $\hat\sigma_x^S$ be performed so as to fulfil the non-invasiveness requirements) for $\theta$ close to $k\pi/2$ (with $k\in\mathbb{Z}$).

A different picture is obtained when the interaction between $S$ and $E$ is sufficiently weak and the state of the latter sufficiently close to an eigenstate of $\hat\sigma^E_z$ to guarantee Markovianity of the system's evolution. In line with the independence of $L_Q(t)$ from $\theta$ mentioned above, although a fully divisible reduced dynamics of the system is guaranteed (as signalled by $\gamma+g(t)\ge0$ and $\sigma(t)\le0$ at all times of the evolution), the extended LG inequality remains violated without changes with respect to Fig.~\ref{centralone} {\bf (a)}. Needless to say, non-Markovianity can no longer be linked to such a result, which must only be due to the occurrence of strong quantum coherence in the state of the system, unaffected by the (weak) interaction with $E$~\cite{theoryJie}. We have verified that, indeed, that under the assumption of non-invasive measurements and for $\theta=\pi/18$, $J=30$Hz, the original LG inequality would be violated, as shown in the Appendix. 

We can gain additional insight in the behaviours shown experimentally in our analysis so far by addressing the formulation of LG arguments given in Refs.~\cite{SusanaSantos,SusanaSantos2}. 
There it was shown that testable LG-type inequalities can be obtained when assuming stationarity for the two-time correlation functions. This avoids the explicit assumption of non invasive measurability at the expense of implicitly assuming the evolution to be Markovian~\cite{EmaryReview}. The evolution induced by this type of classical stochastic process constraints the conditional probability $P(\zeta,t_1|\zeta,t_0)$ to find the system in $\ket{\zeta}$ at time $t_1$ given that it was prepared in such state at time $t_0$ according to the set of inequivalent inequalities
\begin{equation}
\label{susanaineq}
\begin{aligned}
H_1(t)&=P(\zeta,2t|\zeta,0)-P^2(\zeta,t|\zeta,0)\ge0,\\
H_2(t)&=P(\zeta,2t|\zeta,0)+2P(\zeta,t|\zeta,0)\ge1.
\end{aligned}
\end{equation}
Eqs.~\eqref{susanaineq} hold under the additional assumption of time-translational invariance of the conditional probabilities, which translates mathematically into $P(\zeta,t+t_0|\zeta,t_0)=P(\zeta,t|\zeta,0)$, for any $t_0$. Let us notice here that, owing to the way the various assumptions are typically intertwined in the derivation of LG-type inequalities, their violation forces a stochastic classical formalism to incorporate memory effects that mimic the quantum mechanical predictions. This can be seen as the analogous, in the temporal scenario, of the inclusion of non-local effects in hidden variables theories able to reproduce the quantum mechanical correlations of space-like separated subsystems \cite{bohm}. Here, we are interested in the combination of the assumptions of Markovianity and time-translational invariance, whose validity implies the possibility to write classical rate equations for the populations of the set of macroscopic states that describe the system. The violation of Eq.~\eqref{susanaineq} may thus imply the break-down of Markovianity of the system. 

\begin{figure}[t]
\begin{center}
\includegraphics[width=\columnwidth]{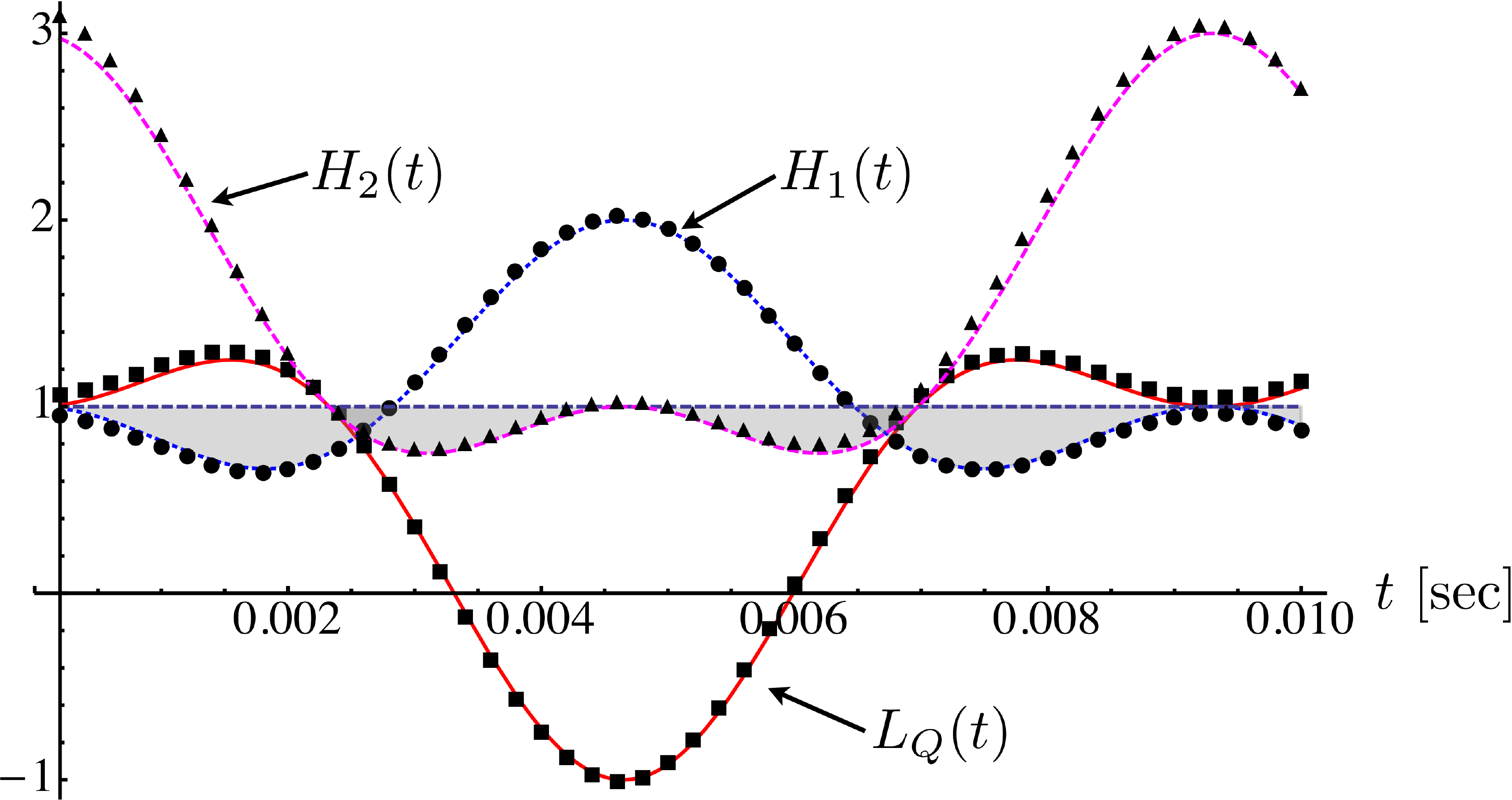}
\end{center}
\caption{{{Time-translational}-based approach to the revelation of non-Markovianity}. Squares, dots and triangles show the experimental values of $L_Q(t)$, $H_1(t)$ and $H_2(t)$, respectively, for the same experimental conditions as in Fig.~2. The solid red, dotted blue, and dashed magenta lines show the corresponding theoretical predictions. The grey-shadowed regions highlight the combined temporal windows of violation of Eq.~\eqref{susanaineq}.}
\label{susana}
\end{figure}

We have embraced this approach in our study and considered experimentally the conditional probability $P(+,t|+,0)$ in regimes of non-Markovianity of the evolution (as addressed above). Quite remarkably, the inequality $H_1(t)\ge0$ implies $|L_Q(t)|\le1$ (cf. Ref.~\cite{SusanaSantos}). However, the second of Eqs.~\eqref{susanaineq} is inequivalent to the others and may thus leads to violations in temporal regions where $H_1(t)$ is strictly positive and Eq.~\eqref{extended} holds. 
In Fig.~\ref{susana} we have found that, indeed, across the whole temporal window explored in our experiment, one or both of Eqs.~\eqref{susanaineq} are violated by the dynamics undergone by $S$,consistently with the non-Markovian nature of the map. In agreement with our expectations, the temporal range within which $H_1(t)<0$ is larger than the one corresponding to a violation of Eq.~\eqref{extended}. 

We have addressed experimentally the dynamics of an open spin system affected by both a Markovian dephasing channel and a simple spin environment that induces strong non-Markovian features. Our experimental assessment is based on the violation of a suitable temporal Bell inequality that has been tested in a two-spin NMR setting. The full experimental control demonstrated over the evolution of the system at hand is sufficient to address non-Markovianity from a wide range of perspectives, ranging from the loss of divisibility of the system's dynamics to the back-flow of information from the spin environment. Our work is, to the best of our knowledge, the first experimental endeavour towards the assessment of temporal Bell-like inequalities as  tools for the revelation of non-Markovian features in quantum evolutions, when subjected to proper caveats. These results open up interesting perspectives for the effective, non-tomographic  characterization of dynamical evolution by combining tools of different nature. The establishment of a connection between the violation of macro realistic inequalities and non-Markovianity is tantalising. For instance, in the case of falsification of one of the inequalities addressed in our work, one could think to use independent information gathered on the non-Markovian nature of the evolution itself, much along the spirit of the investigation reported here, to pinpoint the role that quantum coherences have in such violation. \\

\noindent
\acknowledgments
We thank L. Mazzola, K. Modi, M. G. A. Paris, and B. Vacchini, for insightful discussions on the topics of this paper and C. Emary for enlightening discussions over time on the role of the stationarity assumption in the derivation of LG inequalities. F.L.S. and M.P. are grateful to C. A. Kamienski and K. W. Capelle for the provision of facilities at UFABC for the development of part of this work. D.O.S.P. thanks E. R. deAzevedo and T. J. Bonagamba for discussions. D.O.S.P., F.S., and M.P. are grateful to the Centro Brasileiro de Pesquisas F\'isicas for hospitality during the completion of this work. M.P. and F.L.S. are supported by the CNPq ``Ci\^{e}ncia sen Fronteiras'' programme through the ``Pesquisador Visitante Especial'' initiative (grant nr. 401265/2012-9). S.F.H. acknowledges support from the European Commission through the Collaborative Project PAPETS. M.P. thanks the UK EPSRC (EP/G004579/1), the Alexander von Humboldt Stiftung, and the John Templeton Foundation (Grant ID 43467) for financial support. A.M.S., D.O.S.P., I.S.O., F.L.S., and R.S.S. are members of the Brazilian National Institute of Science and Technology of Quantum Information (INCT/IQ). F.L.S. acknowledges partial support from CNPq (grant nr. 308948/2011-4). 

\section*{APPENDIX}

\renewcommand{\theequation}{A-\arabic{equation}}
\setcounter{equation}{0}  

In this Appendix we provide further information on the experimental setup used for the demonstrated non-Markovian dynamics and 
address quantitatively the case of system's parameters such that a transition to Markovianity is induced by the increased influence of the natural dephasing mechanism 
affecting the system spin.

\subsection{Derivation of the time-local map}
{
We present the derivation of the general map and time-local master equation describing the reduced dynamics of the system $S$. We will follow the approach discussed in Refs.~\cite{anderson,smirne}. The map responsible for the evolution of the system, which we write formally as $\rho_S(t)=\Phi[\rho_S(0)]$, is given in Eq.~\eqref{channel} and can be represented as a matrix $M_{\Phi}$ acting on the vector $\varrho_S=\left(\rho_{00}\,\rho_{01}\,\rho_{10}\,\rho_{11}\right)^T$ whose elements are the entries of the density matrix $\rho_{S}(t)$. We can decomposed such matrix using any suitable complete basis $\{\chi_i\}_{i=0,\ldots,4}$ as \cite{smirne}
\begin{equation}\label{matrixmap}
[M_\Phi]_{ij}={\rm{Tr}}[\chi_{i}\Phi[\chi_{j}]],
\end{equation}
where $\Phi[\chi_{j}]$ is obtained using Eq.~(\ref{channel}) with $\chi_{j}$ instead of $\rho_S(0)$. Here, it is convenient to choose $\chi_0=\id/\sqrt{2},\chi_i=\sigma_i/\sqrt{2}$, where $\{\sigma_i\}_{i=1,2,3}$ is the set of Pauli matrices. Clearly, $M_\Phi(t)$ is \textsl{uniquely} determined by the map (it does not depend on the initial system state $\rho_S(0)$). In order to express the time-local master equation in the form $\partial_t{\rho}_S(t)=\hat{\cal{K}}(t)[\rho_S(t)]$ (with $\hat{\cal K}(t)$ a time-local superoperator), we call $K(t)$ the matrix associated with ${\cal{K}}(t)$ and write 
\begin{eqnarray}\label{K}
K(t)=(\partial_tM_\Phi)M_\Phi^{-1}.
\end{eqnarray} 
In our case, the inverse matrix $M_\Phi^{-1}$ is straightforwardly evaluated and we obtain 
\begin{equation}
\label{Kmatrix}
K(t)=
\begin{pmatrix}
0 & 0 & 0 & 0 \\
 0 & g(t) & -f(t) & 0\\
 0 & f(t) & g(t) & 0\\
 0 & 0 & 0 & 0
\end{pmatrix}.
\end{equation}
By decomposing this matrix in the chosen basis, it is immediate to find
\begin{equation}
\partial_t\rho_{S}(t)=if(t)[\hat\sigma_z^S,\rho_S(t)]+g(t)[\hat\sigma_z^S\rho_S(t)\hat\sigma^S_z-\rho_S(t)].
\end{equation}
When the system's dephasing is included in the description of the evolution, the same procedure outlined above can be applied, leading directly to Eq.~(\ref{ME}).

\subsection{Further details on the experimental setup}

The experiments were performed on a $^{13}$C-enriched chloroform sample (CHCl3). The two-level systems used in our work have been encoded in the $^1$H and $^{13}$C nuclear spins. The sample was prepared by mixing 100 mg of 99\% $^{13}$C-labelled CHCl$_3$ in 0.7 mL of 99.8\% CDCl$_3$ in a 5 mm NMR tube with both compounds provided by the Cambridge Isotope Laboratories Inc. The NMR experiments were carried out at 25$^{\circ}$C in a Varian 500 MHz Premium Shielded spectrometer located at the Brazilian Center for Research in Physics (CBPF, Rio de Janeiro) using a Varian 5 mm double resonance probehead equipped with a magnetic field gradient coil. The spin-lattice (spin-spin) relaxation times $T_1$ for the $^1$H and $^{13}$C nuclei, measured by the inversion-recovery pulse sequence (CPMG pulse sequence), were $3.57$s and $10$s ($1.2$s and $0.19$s), respectively. 
The recycle delay was set at $90$s in all experiments. 

The nuclear spin Hamiltonian is given by
\begin{equation}
\label{lab}
\begin{aligned}
\hat{\cal H} &=- (\omega_{H}-\omega_{rf}^{H})\,\hat{I}_{z}^{H} - (\omega_{C}-\omega_{rf}^{C})\,\hat{I}_{z}^{C} + 2\pi\,J\,I_{z}^{H}\,\hat{I}_{z}^{C}\\
&+\omega_{1}^{H}\,(\hat{I}_{x}^{H}\cos\phi^{H}+\hat{I}_{y}^{H}\sin\phi^{H}) + \omega_{1}^{C}\,(\hat{I}_{x}^{C}\cos\phi^{C}+\hat{I}_{y}^{C}\sin\phi^{C}),
\end{aligned}
\end{equation}
where $\hat{I}_{\alpha}^{H}$ ($\hat{I}_{\beta}^{C}$) is the spin angular momentum operator in the $\alpha, \beta = x, y, z$ direction for $^1$H ($^{13}$C), $\phi^{H}$ ($\phi^{C}$) defines the direction of the rf field (pulse phase) and $\omega^{H}_{rf}$ ($\omega^{C}_{rf}$) is the rf nutation frequency (rf power) for the $^1$H ($^{13}$C) nuclei. Eq.~\eqref{lab} is written in the rotating frame. In this case the rf terms are time-independent and the terms of the form $\Delta\omega^a = \omega - \omega^a_{rf}~(a=H,C)$ represent the frequency offset of each nucleus.

The first two terms in Eq.~\eqref{lab} describe the Zeeman interaction between the $^1$H and $^{13}$C nuclear spins and the main magnetic field $B_0$. The corresponding frequencies are $\omega_H / 2\pi \approx$ 500 MHz and $\omega_C / 2\pi \approx$ 125 MHz. The third term is due to a scalar spin-spin coupling having coupling rate $J \approx$ 215 Hz. The fourth and fifth terms describe the precession induced by the rf field applied to the $^1$H and $^{13}$C nuclear spins, respectively. A time-dependent coupling of the nuclear spins with the environment that includes all fluctuating NMR interactions (such as $^1$H-$^{13}$C dipolar spin-spin couplings and interactions with the chlorine nuclei) and accounting for the spin relaxation has also been considered in our treatment of the dynamics.

\subsection{Quality of the experimental data}

In this Section we address the quality of the data that we have measured in our experiments by addressing some significant figures of merit that show the high consistency between the data and our theoretical predictions.

We start assessing the closeness of the reconstructed $S$-$E$ states to the expected ones. To this aim we use states fidelity as defined in Ref.~\cite{fidelityNMR} and customarily used in NMR experiments 
\begin{equation}
{\cal F}=\frac{|{\rm Tr}(\rho_0\rho_1)|}{\sqrt{{\rm Tr}(\rho^2_0){\rm Tr}(\rho_1^2)}}
\end{equation}
with $\rho_{0,1}$ two density matrices. Using experimental two-qubit quantum state tomography in NMR systems~\cite{tomoNMR}, we have reconstructed the density matrix of the system and environment spins at set values of the parameters $\theta,\,J$ and $\omega$ defining the evolution at hand. Fig.~\ref{tomo&fide} shows the fidelity between theoretical and experimental density matrices corresponding to the data shown in Fig.~2 of the main text. The high quality of the experimental state is evident with fidelities well above $98.5\%$ across the whole sample of reconstructed states and negligible differences between the two different initial preparations of the $S$ spin. Data of similar quality are found for the various values of $\theta$ (at fixed value of $J$) that we have considered in our experiment, as shown in Fig.~\ref{fide2}. In this context, while conclusions analogous to those reported in the main text regarding the violation of the generalised LG inequality and the tomography of the master-equation hold basically unchanged, it is interesting to discuss further the case of $\theta=\pi/3$, which has been extensively addressed in the main text, and briefly assess the case of $\theta=\pi/4$.

\begin{figure}[t]
\includegraphics[width=\columnwidth]{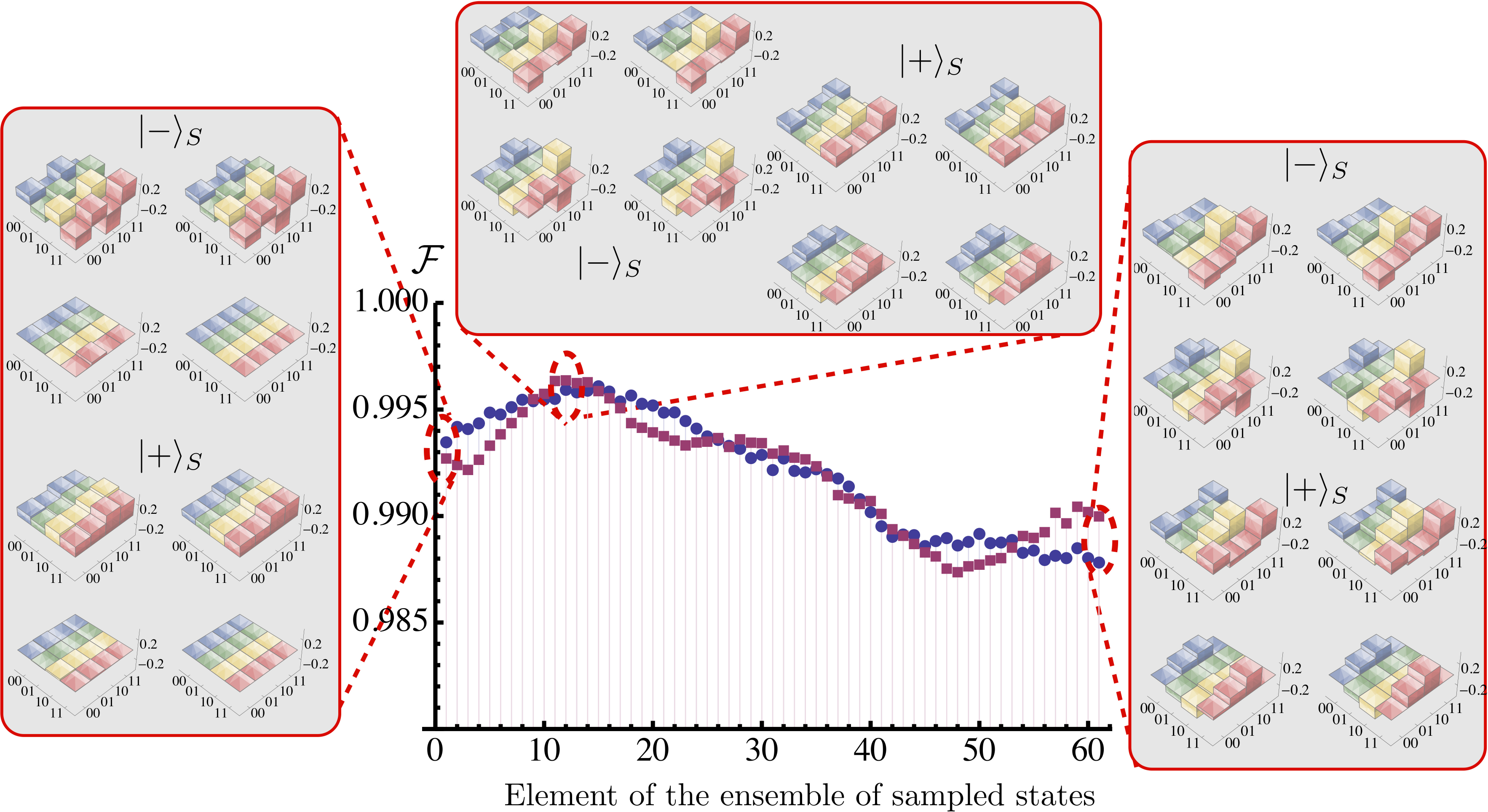}
\caption{State fidelity between the experimentally reconstructed $S$-$E$ states and their expected form. We have taken $J=215.16$Hz with $\theta=\pi/3$. The main panel shows the distribution of the values of the fidelity ${\cal F}$ sampled in steps of $0.2$ms from time $t_i=0$ to the $t_f=12$ms of the evolution. Blue dots (Purple squares) show the fidelity corresponding to the system being initialised in $\ket{+}_S$ ($\ket{-}_S$). Each inset shows the bar chart tomographies of selected states of the sample. We show the cases corresponding to $t=0$ (leftmost inset), $t=2.2$ms (central inset), and $t=12$ms (rightmost one). In each inset, we show data corresponding to both preparations of the $S$ spin. The left (right) column of each inset shows the experimental (theoretical) density matrices: for each preparation of $S$, the top (bottom) row shows the real (imaginary) part of the density matrix entries.}
\label{tomo&fide}
\end{figure}

\begin{figure}[b!]
{\bf (a)}\hskip4cm{\bf (b)}\\
\includegraphics[width=\columnwidth]{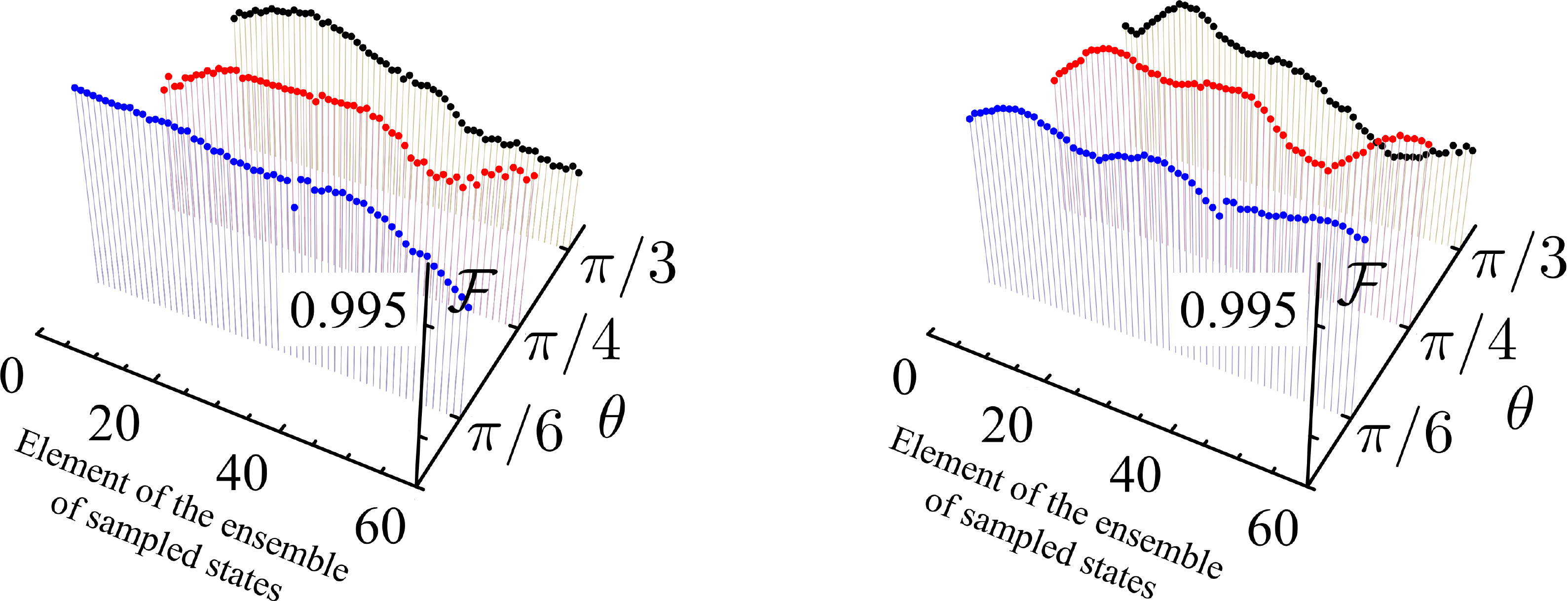}
\caption{Distribution of the values of quantum state fidelities between the theoretical and experimental $S$-$E$ states corresponding to $J=216.06$Hz and $\theta=\pi/6,\pi/4,\pi/3$. We have studied the evolution up to $t=12$ms by sampling the $S$-$E$ states at steps of $0.2$ms. Panel {\bf (a)} and {\bf (b)} show the fidelities corresponding to the preparations $\ket{+}_S$ and $\ket{-}_S$ respectively.}
\label{fide2}
\end{figure}

\begin{figure}[b!]
{\bf (a)}\hskip4cm{\bf (b)}
\includegraphics[width=.5\columnwidth]{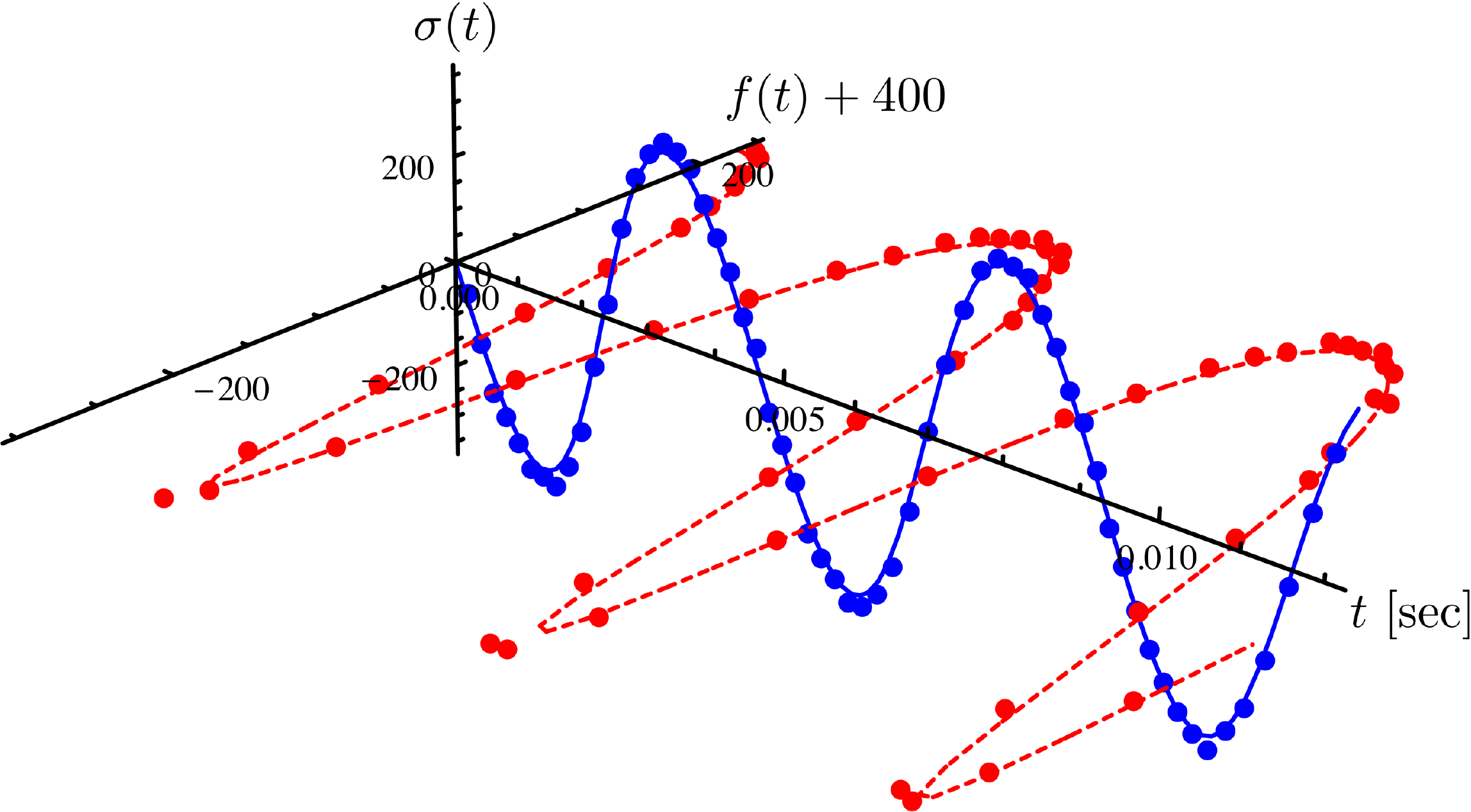}\includegraphics[width=0.5\columnwidth]{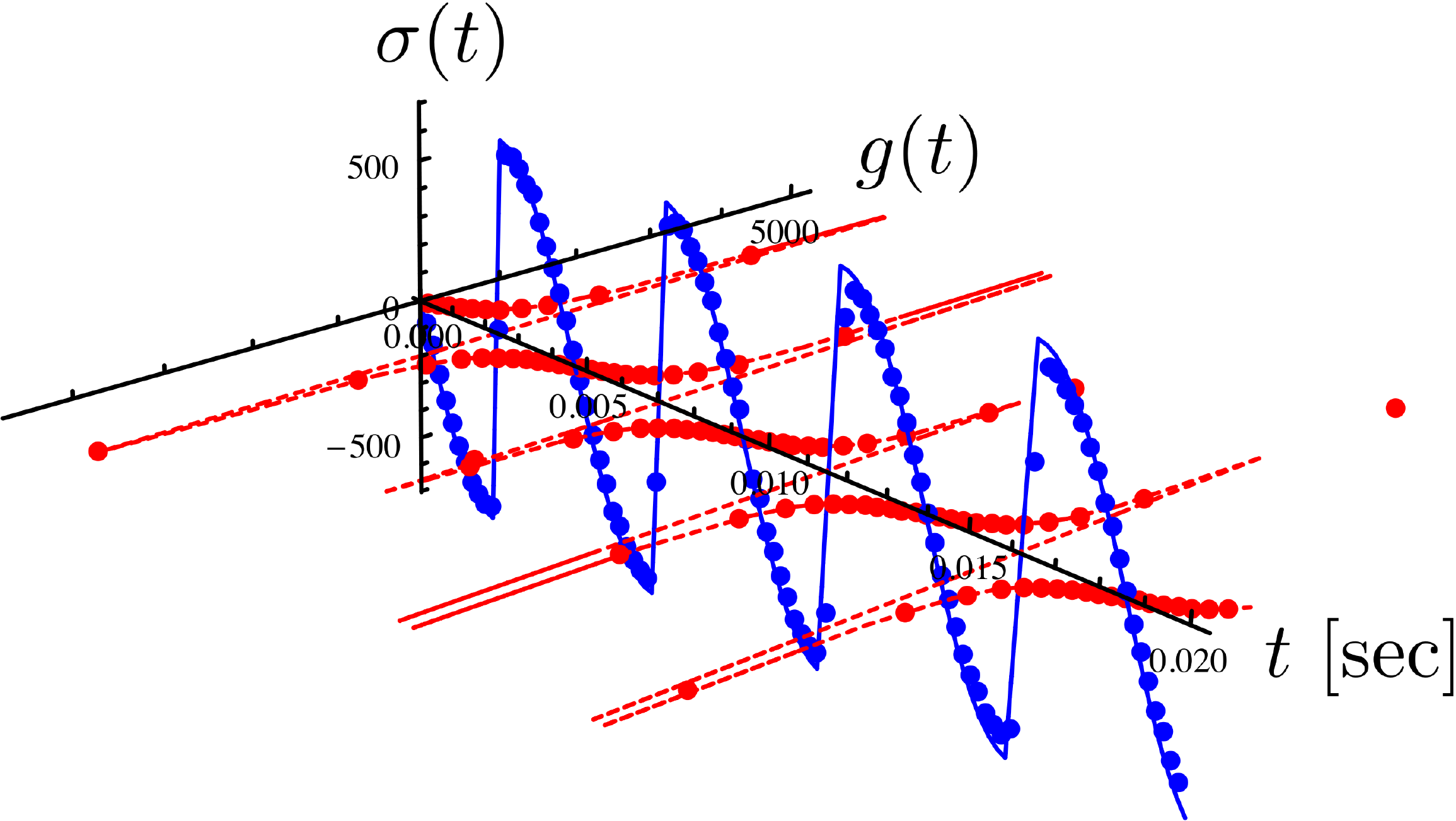}
\caption{{\bf (a)} Temporal behaviour of $f(t)$ and $\sigma(t)$ for $J=215.06$Hz, $\theta=\pi/3$ and $S$ prepared in $\ket{-}_S$. For convenience of comparison, we have shifted $f(t)$ by a uniform amount. When $f(t)$ reaches a minimum, $\sigma(t)$ changes sign, thus marking the onset of non-Markovianity. {\bf (b)} Temporal behaviour of $g(t)$ and $\sigma(t)$ for $J=215.06$Hz, $\theta=\pi/4$ and $S$ prepared in $\ket{-}_S$. At times such that $Jt=\pi/2+k\pi$, the $g(t)$ function diverges, marking the change in sign of $\sigma(t)$, and thus the onset of non-Markovianity. A similar behaviour of $g(t)$ is observed when $S$ prepared in $\ket{+}_S$. In both panels, dashed and solid lines are theoretical expectations, while dots show the experimental data. }
\label{ThetaPiOver4}
\end{figure}

We complement the analysis reported in the main text regarding the case $\theta=\pi/3$ by discussing briefly the relation between the function $f(t)$ entering the reduced master equation for the $S$ dynamics and $\sigma(t)$. In Fig.~\ref{ThetaPiOver4} {\bf (a)}, we plot both such functions against the evolution time $t$, showing that the onset of non-Markovianity, as given by the change of sign of $\sigma(t)$ from negative to positive values, occur when $f(t)$ achieves its minima.

For $\theta=\pi/4$, the time-dependent coefficients entering Eq.~(5) of the main text give have singularities at $\pi Jt=\pi/2+k\pi~(k\in{\mathbb Z})$. Correspondingly, the quantity $\sigma(t)$ that has been related to the measure of non-Markovianity based on the trace distance turns positive, therefore marking the non-Markovian trend of the dynamical evolution of $S$. It is thus interesting to check that the experimental data are able to reveal such singular behaviour, as shown in Fig.~\ref{ThetaPiOver4}, where a perfect correlation between the divergence of the experimental values of $g(t)$ and the change of sign from negative to positive of the inferred $\sigma(t)$ is clearly observed. Notice the considerable growth of the amplitude of oscillation of $\sigma(t)$ with respect to what is observed in the main text for $\theta=\pi/3$.

As a final consistency test, we assess the relation between the transverse magnetisation and $\sigma(t)$, thus linking an observable that is easily accessed to the occurrence of non-Markovian 
features. In order to do this, we consider the values of $\sigma(t)$ and $\langle\hat\sigma_-(t)\rangle$ (both real and imaginary part) corresponding to a given value of $t$, and plot them against each other for the preparation corresponding to $\ket{-}_S$ and $\theta=\pi/3$ in Fig.~\ref{farfalle}. It is interesting to notice that, both being oscillating functions of time, $\sigma(t)$ has twice the frequency of ${\rm Re}[\langle\hat\sigma_-(t)\rangle]$ and ${\rm Im}[\langle\hat\sigma_-(t)\rangle]$. Only when the real (imaginary) part of the magnetisation achieves zero (its maximum value), $\sigma(t)$ turns from negative to positive values (from the expression of $\eta_\theta(t)$ given in the main text, it is straightforward to see that the real and imaginary part of the magnetisation are out of phase by $\pi/2$).

\begin{figure}[t]
\hskip-1cm{{\bf (a)}\hskip4cm{\bf (b)}}
\includegraphics[width=\columnwidth]{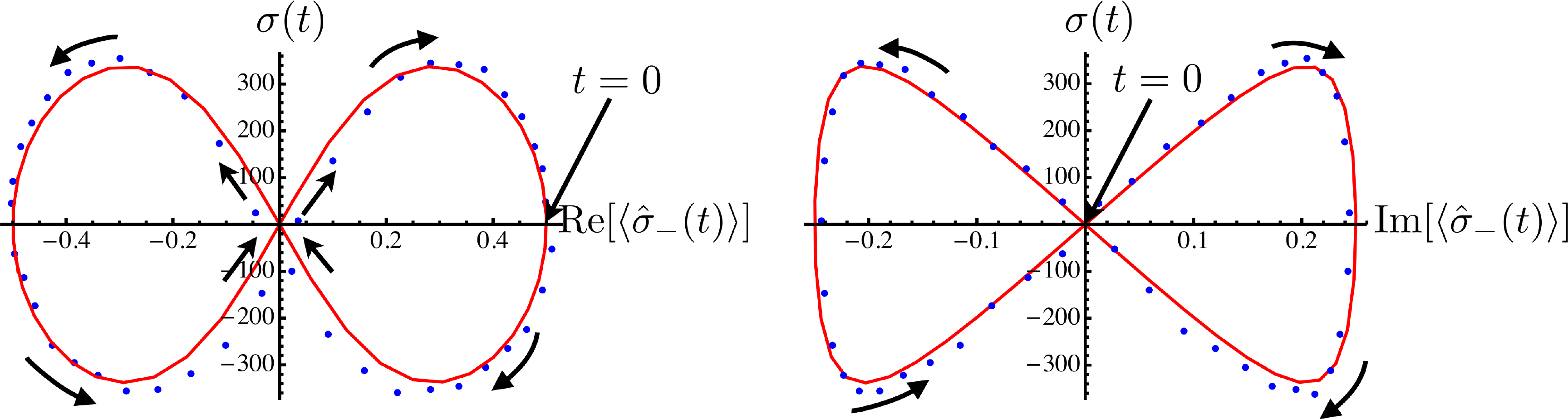}
\caption{Comparison between the transverse magnetisation and the function $\sigma(t)$ evaluated at the same instant of time (which is a curvilinear abscissa). We assess individually both the real and the imaginary part of $\langle\hat\sigma_-(t)\rangle$. The arrows show the way time grows in the evolution. Panel {\bf (a)} [{\bf (b)}] shows $\sigma(t)$ against the real (imaginary part) of the transverse magnetisation.}
\label{farfalle}
\end{figure}

\subsection{Transition to Markovianity}

In this Section we address the case of experimental conditions such that the dephasing mechanism at rate $\gamma$ in Eq.~(5) of the main text is not negligible with respect to the $S$-$E$ interaction. In our experiment we have found $\gamma^{-1}\simeq150\times10^{-3}$s measured by the so-called CPMG technique~\cite{refDiogo}. In order to make the dynamics of the system divisible, and thus Markovian, we need to ensure that the coefficients of the master equation Eq.~(5) are positive, thus certifying divisibility of the corresponding map. At set values of $J$ and $\gamma$ (which are determined by the experimental conditions), this can be enforced by properly preparing the environment spin, i.e. by choosing $\theta$ judiciously. In fact, if the environment is prepared in an eigenstate of $\hat\sigma^E_z$, the $S$-$E$ coupling is effectively turned off, thus leaving the system to the sole effect of the dephasing mechanism. One thus expect that, for $\theta$ sufficiently low, the system-environment interaction would not be able to overcome the Markovianity induced by the natural dephasing of the system. Such threshold in $\theta$ can be calculated by taking A Taylor expansion of $g(t)$ in Eq.~(6) of the main text and comparing it with the dephasing rate $\gamma/2$ in the system's time-local master equation. This leads us to the condition $\theta\in[-\theta_M,\theta_M]$ with
\begin{equation}
\theta_M=\sqrt{\frac{\gamma\csc(2\pi J t)}{2\pi J}}.
\end{equation}
A typical behavior of $\theta_M$ for a small value of the coupling coefficient $J$ is shown in Fig.~\ref{angolo}. A choice of $\theta=\pi/18$, as in our experiment, guarantees that the coefficient of the reduced master equation are do not change sign as $g(t)<\gamma/2$ at all instants of time, in this case. This can be seen from Fig.~\ref{centralone1} {\bf (b)}, where we plot $\tilde{g}(t)=\gamma/2+g(t)$ for $J=30$Hz, $\theta=\pi/18$, $\gamma^{-1}=150$ms.
\begin{figure}[t]
\includegraphics[width=\columnwidth]{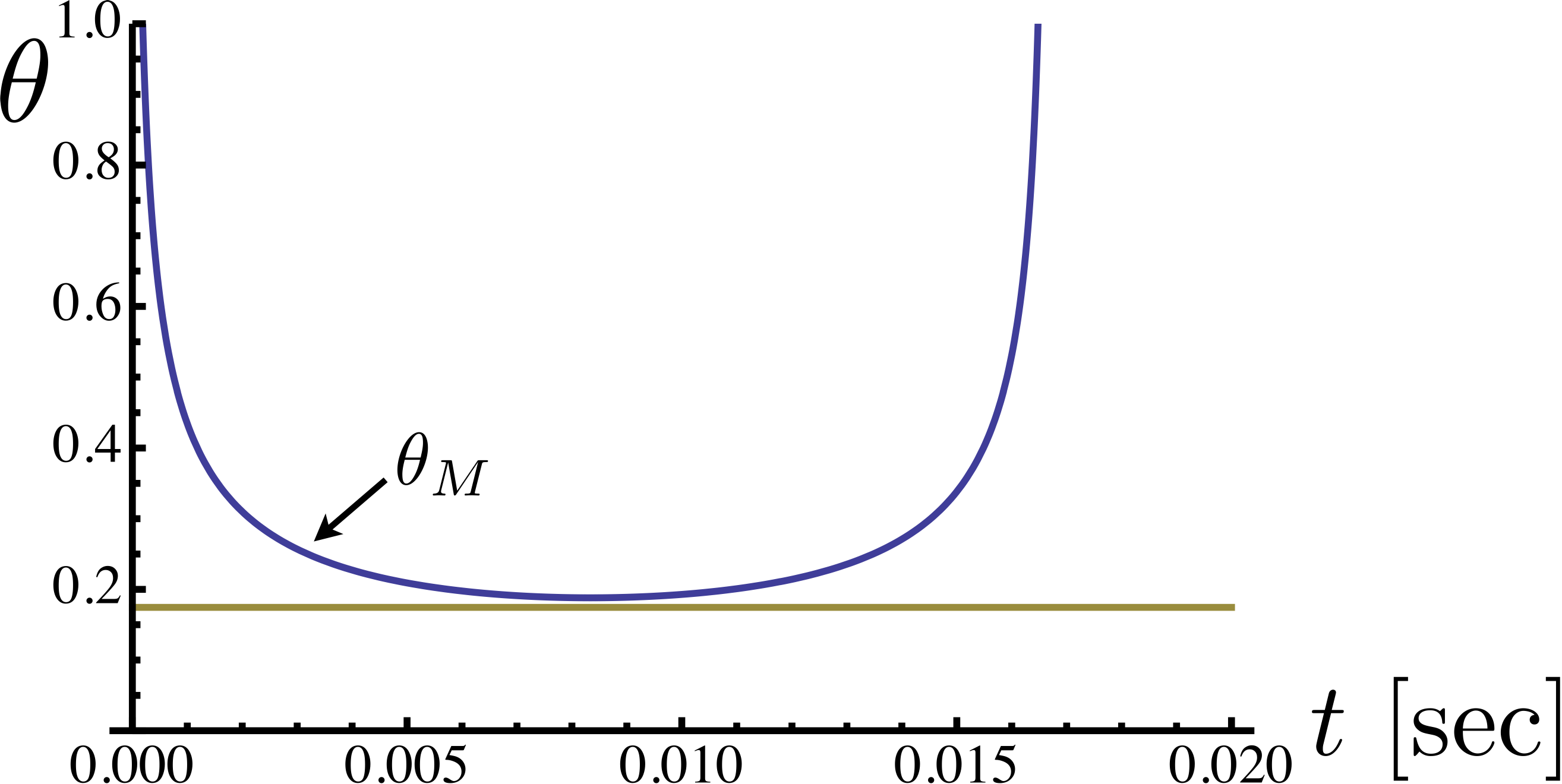}
\caption{Temporal behaviour of the threshold value $\theta_M$ of the angle $\theta$ such that, at set values of $J$, the dynamics of $S$ is Markovian. We have explored only the region $\theta\in[0,\theta_M]$. The straight line shows $\theta=\pi/18$, as in the experiment reported in the main text, which is thus well within the Markovianity region. }
\label{angolo}
\end{figure}

\begin{figure}[b]
\begin{center}
{\bf (a)}\hskip2.5cm{\bf (b)}\hskip2.5cm{\bf (c)}
\includegraphics[width=0.33\columnwidth]{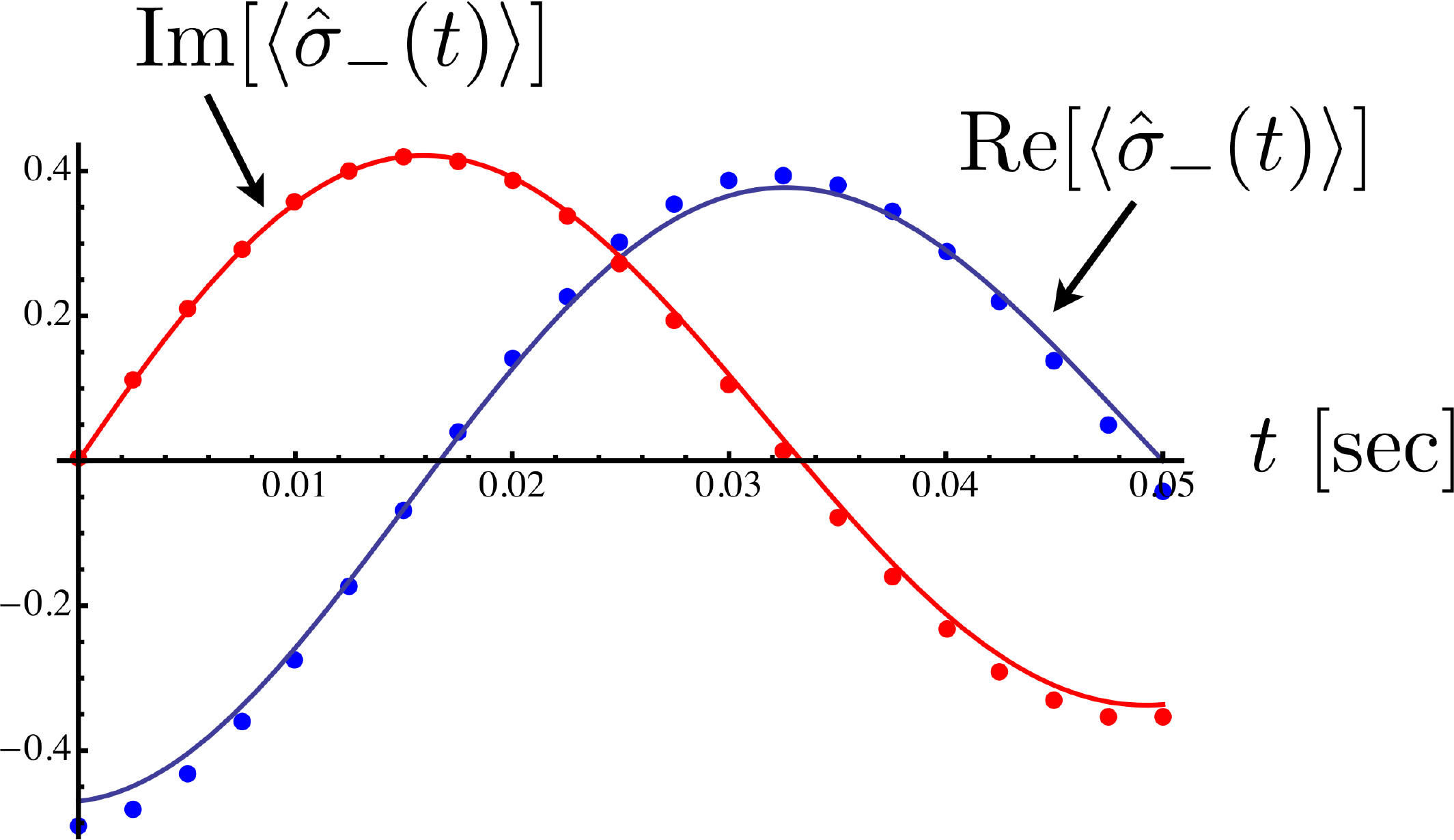}\includegraphics[width=0.33\columnwidth]{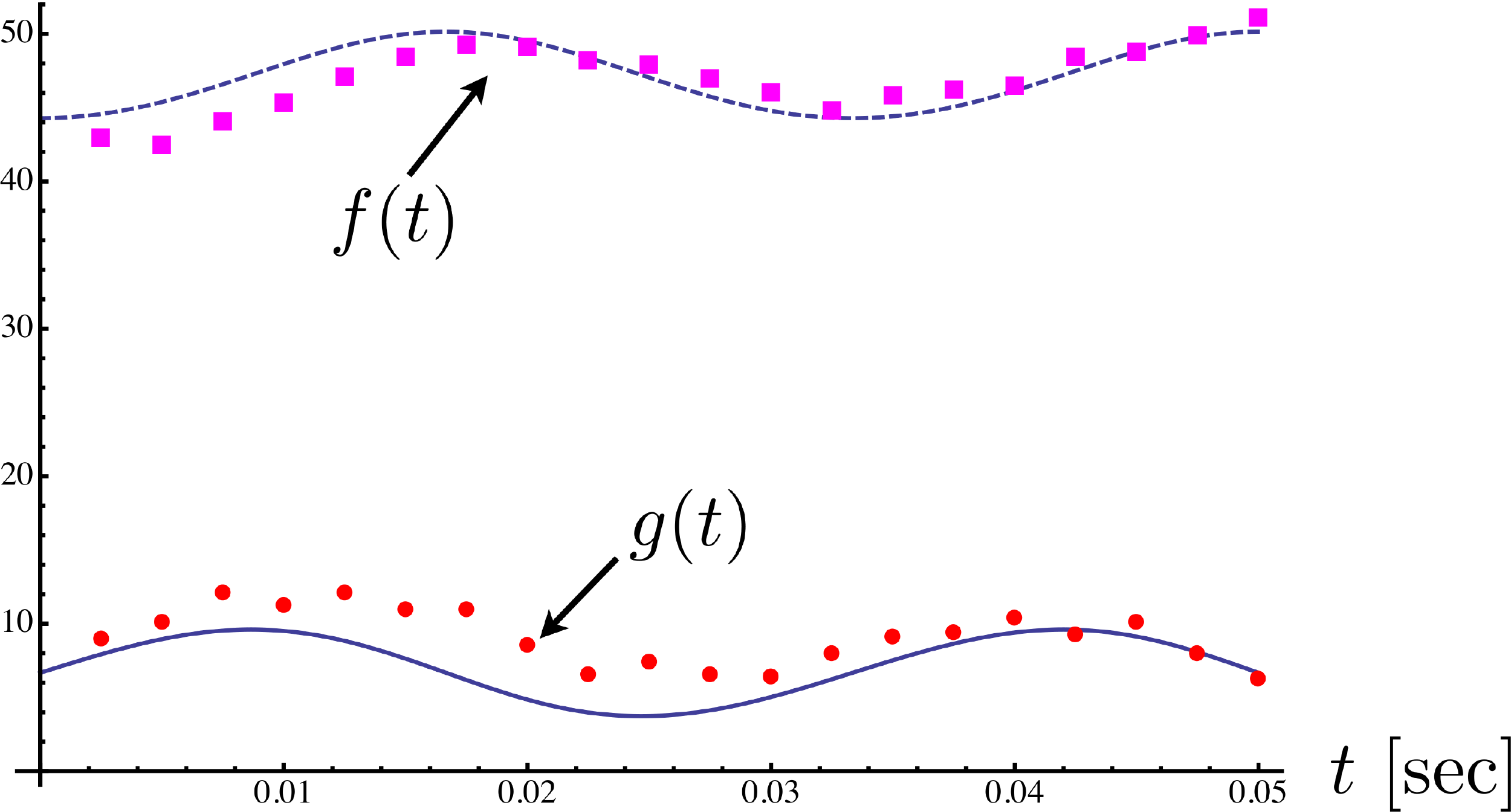}\includegraphics[width=0.35\columnwidth]{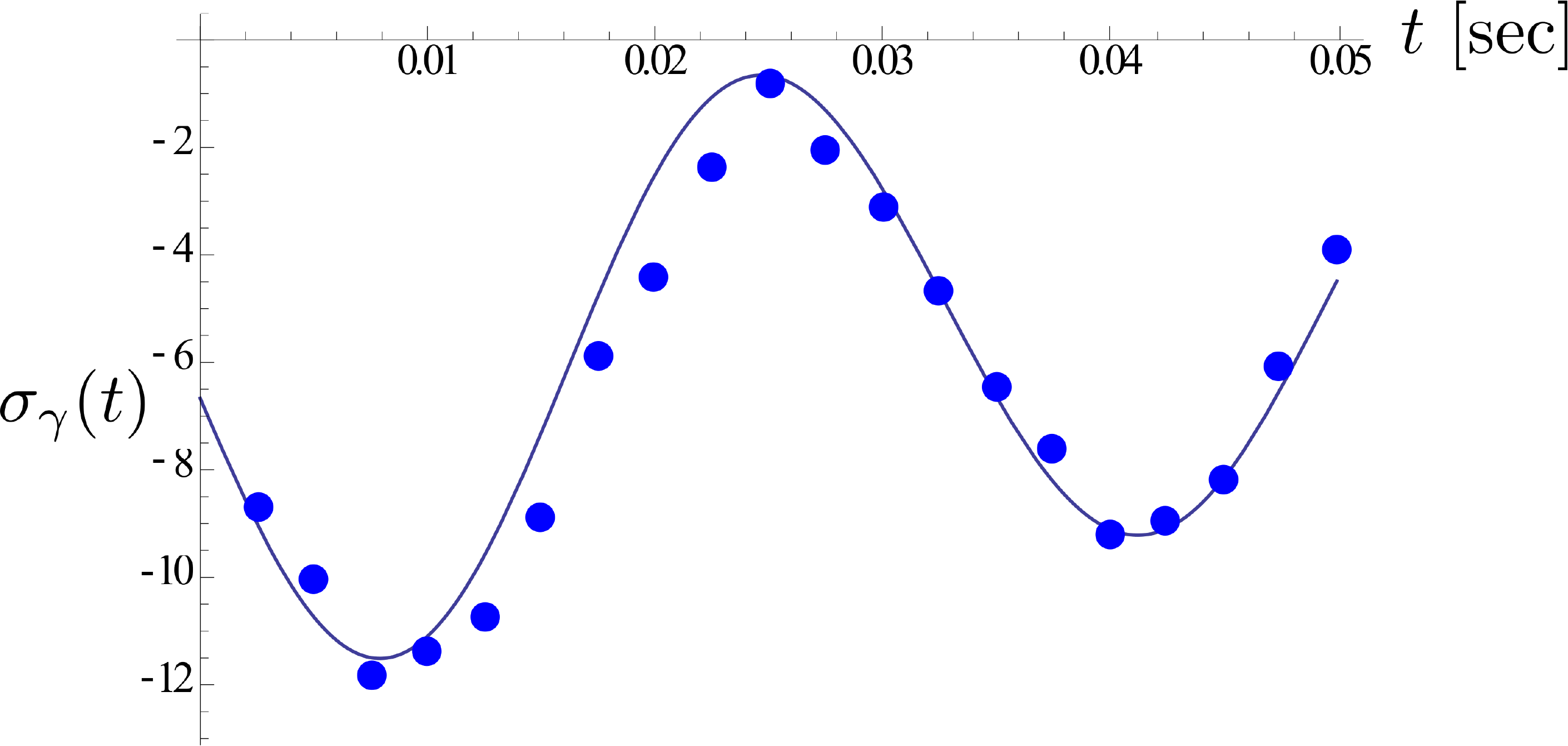}
\end{center}
\caption{{\bf (a)} Real and imaginary part of the transverse magnetisation of the system $S$ for an initial preparation $\ket{-}_S$, $J=30$Hz, $\theta=\pi/18$, and $1/\gamma=150$ms. {\bf (b)} Reconstructed functions $f(t)$ and $\tilde{g}(t)$ for the same configuration as in panel {\bf (a)}. Points represent experimental data, while the solid and dashed lines are the theoretical predictions. {\bf (c)} Dephasing-affected $\sigma_\gamma(t)$ function. Its strictly non-positive values witness that no back-flow of information from the environment influences the state of the system, in line with the divisibility of the reduced dynamical map for $S$.}
\label{centralone1}
\end{figure}

\begin{figure}[t]
{\bf (a)}\hskip4cm{\bf (b)}
\includegraphics[width=.5\columnwidth]{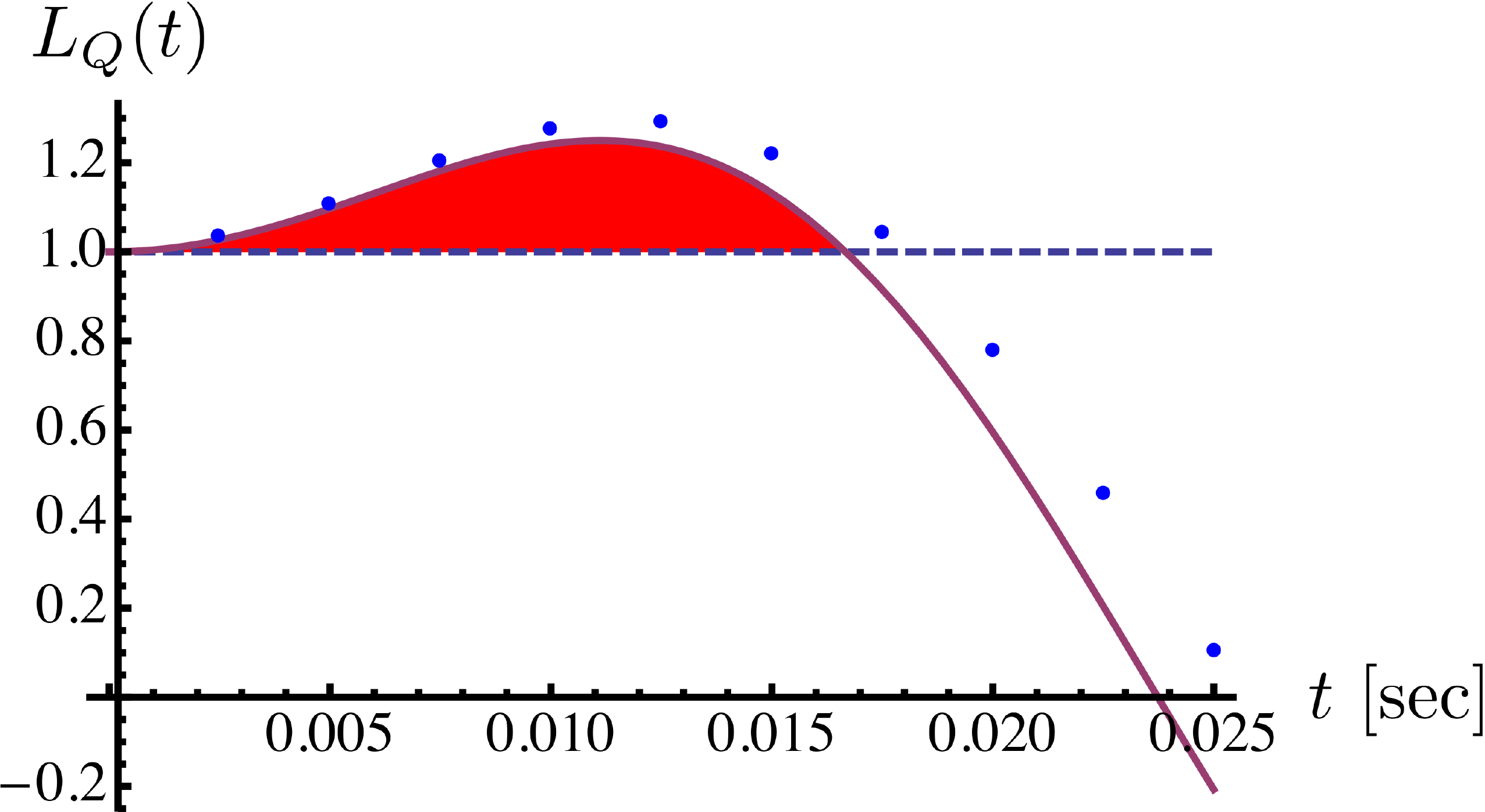}~\includegraphics[width=.5\columnwidth]{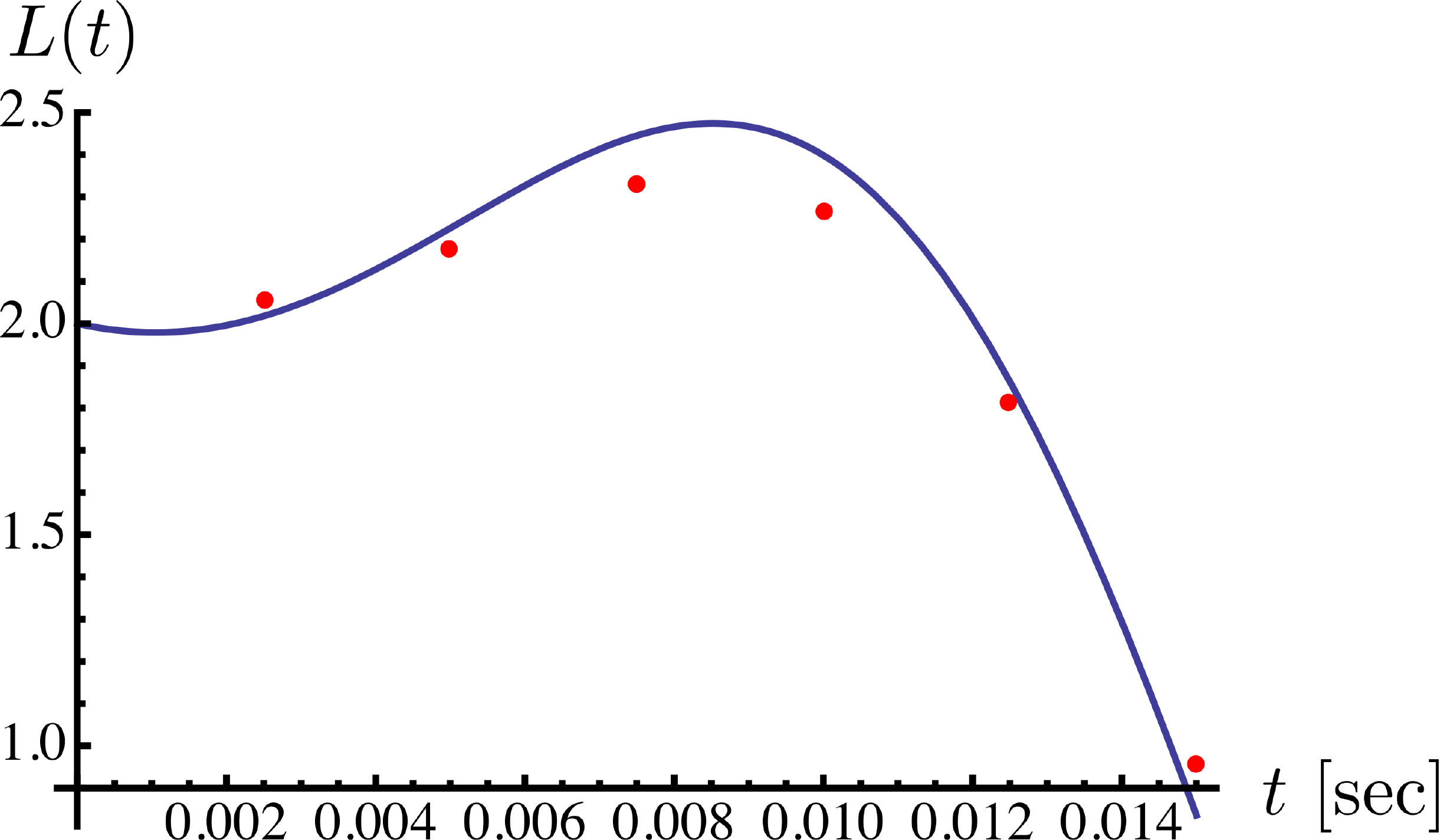}
\caption{{\bf (a)} Violation of the generalised LG inequality by the state of the system $S$ for $J=30$Hz, $\theta=\pi/18$, $1/\gamma=150$ms and the initial preparation $\ket{+}_S$. The solid line (circles) show the behavior of the theoretical $L_Q(t)$ function (experimental data), while the shadow identifies the region of violation of classical Markovian assumptions. {\bf (b)} Violation of the standard LG inequality}
\label{violazioni}
\end{figure}

The divisibility of the associated dynamical map can then be confirmed by the negativity of the quantity $\sigma(t)=\partial_t||\rho_{S,1}(t)-\rho_{S,2}(t)||$ used in the main text. This figure of merit, though, should be reformulated in order to take into account the effects of dephasing. Such reformulation is actually straightforward and leads us to  
\begin{widetext}
\begin{equation}
\sigma_\gamma(t)=-{e^{- \gamma  t}}\frac{\gamma  [\cos (4 \theta )+3]+2 \sin ^2(2 \theta ) [\gamma  \cos (2 \pi  J t)+\pi  J \sin (2 \pi  J t)]}{2
   \sqrt{2 \cos (4 \theta ) \sin ^2(\pi  J t)+\cos (2 \pi  J t)+3}}.
\end{equation}
\end{widetext}
In Fig.~\ref{centralone1} (c) we have compared the behavior of $\sigma_{\gamma}(t)$ to the trend that we have inferred experimentally for a small coupling $J=30$Hz and angle $\theta=\pi/18$. As anticipated in the main text, the demonstrated Markovianity of the system's dynamics does not hinders the violation of the generalised LG inequality. Indeed, by following an approach similar to the one described in the main text but for the dephasing-affected correlation function, we get the behaviour shown in fig.~\ref{violazioni} {\bf (a)}. Yet, the violation of the generalised LG inequality, in this case, cannot be ascribed to any non-Markovianity, but to the non-classical nature of the evolution of $S$. This can be demonstrated by assuming the possibility to measure the observable $\hat\sigma_x^S$ in a non-invasive way (which is not possible with the setup used in our experiment) and addressing the standard LG inequality~\cite{kofler}
\begin{equation}
L(t)={\cal C}(t;0)+{\cal C}(2t;t)+{\cal C}(3t;2t)-{\cal C}(3t;0)\le2
\end{equation}
with ${\cal C}(t_2;t_1)=\langle\hat\sigma^S_{x}(t_2)\hat\sigma^S_{x}(t_1)\rangle$ and $t_2>t_1$. The violation of this inequality implies the untenability of the assumptions of realism per se and non-invasive measurements~\cite{leggett}, which are well-accepted features of a fully classical theory. Fig.~\ref{violazioni} {\bf (b)} shows that, should the measurement of $\hat\sigma_x^S$ be performed non-invasively, the evolution of the system's spin would be certified as non-classical, hence the falsification of the generalised inequality.



\begin{thebibliography}{99}
%
\bibitem{petruccione} H.-P. Breuer and F. Petruccione, {\it The Theory of Open Quantum Systems} (Oxford University Press, Oxford  2002).

\bibitem{RivasHuelga} \'A. Rivas, and S. F. Huelga, {\it Open Quantum Systems. An Introduction} (Springer Briefs in Physics, Springer 2011).

\bibitem{applications} E.-M. Laine, H.-P. Breuer, and J. Piilo, arXiv:1210.8266 (2012); 

\bibitem{applications2} S. McEndoo, P. Haikka, G. De Chiara, M. Palma, S. Maniscalco, EPL {\bf 101}, 60005 (2013).


\bibitem{applications4}S. F. Huelga, \'A. Rivas, and M. B. Plenio, Phys. Rev. Lett. {\bf 108}, 160402 (2012); 

\bibitem{applications5} A. W. Chin, S. F. Huelga, and M. B. Plenio, Phys. Rev. Lett. {\bf 109}, 233601 (2012).

\bibitem{applications6} R. Vasile, S. Olivares, M. G. A. Paris, and S. Maniscalco, Phys. Rev. A 83, 042321 (2011).

\bibitem{SusanaNatPhys} A. W. Chin, J. Prior, R. Rosenbach, F. Caycedo-Soler, S. F. Huelga, and M. B. Plenio, Nature Phys. {\bf 9}, 113 (2013).
 
\bibitem{Aspuru} P. Rebentrost, R. Chakraborty,and A. Aspuru-Guzik, J. Chem. Phys. {\bf 131}, 184102 (2009). 
 
\bibitem{Li} C.-M. Li, N. Lambert, Y.-N. Chen, G.-Y. Chen, and F. Nori, Sci. Rep. {\bf 2}, 885 (2012); N. Lambert, Y.N. Chen, Y.C. Chen, C.M. Li, G.Y. Chen, F. Nori, Nature Phys. {\bf 9}, 10 (2013).

\bibitem{nori_lambert1} N. Lambert, C. Emary, Y. -N. Chen, and F. Nori, Phys. Rev. Lett \textbf{105}, 176801 (2010).
%
\bibitem{EmaryReview} C. Emary, N. Lambert, and F. Nori,  arXiv:1304.5133.
%
\bibitem{NMRbooks} A. Abragam, \textit{The Principles of Nuclear Magnetism} (Oxford University, New York, 1978); R. R. Ernst, G. Bodenhausen, and A. Wokaum, \textit{Principles of Nuclear Magnetic Resonance in One and Two Dimensions} (Oxford University, New York, 1987).
%
\bibitem{leggett_garg} A. J. Leggett and A. Garg, Phys. Rev. Lett. {\bf 54}, 857 (1985).
%
\bibitem{pnas} M. E. Goggin, M. P. Almeida, M. Barbieri, B. P. Lanyon, J. L.
O' Brien, A. G. White, and G. J. Pryde, Proc. Natl. Acad. Sci.
USA \textbf{108}, 1256 (2011).

\bibitem{Dressel} J. Dressel, C. J. Broadbent, J. C. Howell, and A. N. Jordan, Phys. Rev. Lett. {\bf 106}, 040402 (2011).

%
\bibitem{lg_cbpf} A. M. Souza, I. S. Oliveira, and R S Sarthour, New J. Phys. \textbf{13}, 053023 (2011).
%
\bibitem{Athalyse}  V. Athalye, S. S. Roy, and T. S. Mahesh, Phys. Rev. Lett. {\bf 107}, 130402 (2011).

%
\bibitem{laloy} A. Palacios-Laloy, F. Mallet, F. Nguyen, P. Bertet, D. Vion,
D. Esteve, and A. N. Korotkov, Nature Phys. \textbf{6}, 442 (2010).

\bibitem{Knee} G. C. Knee {\it et al.}, Nature Commun. {\bf 3}, 606 (2012).
%

\bibitem{Waldherr} G. Waldherr, P. Neumann, S. F. Huelga, F. Jelezko, and J. Wrachtrup, Phys. Rev. Lett. {\bf 107}, 090401 (2011).


%
\bibitem{cbpfbook} I. S. Oliveira, T. J. Bonagamba, R. S. Sarthour, J. C. C.
Freitas and E. R. de Azevedo, \textit{NMR Quantum Information Processing} (Elsevier, Amsterdam,
2007). 
%
\bibitem{VanderChuang} L. M. K. Vandersypen and I. L. Chuang, Rev. Mod. Phys. \textbf{76}, 1037 (2004).
%

\bibitem{amsouzaDD} A. M. Souza, G. A. Alvarez, D. Suter, Phil. Trans. R. Soc. A \textbf{370}, 4748 (2012).

\bibitem{nielsenchuang} M. A. Nielsen, and I.  Chuang, {\it Quantum Computation
and Quantum Information} (Cambridge University Press,
Cambridge, 2000).

\bibitem{anderson} E. Andersson, J. D. Cresser, and M. J. W. Hall, J. Mod. Opt. {\bf 54}, 1695 (2007).
%
\bibitem{smirne} A. Smirne, and B. Vacchini, Phys. Rev. A {\bf 82}, 022110 (2010).
%

\bibitem{Rivas} \'A. Rivas, S. F. Huelga, and M. B. Plenio, Phys. Rev. Lett. {\bf 105}, 050403 (2010).

\bibitem{breuer} H.-P. Breuer, E. M. Laine, and J. Piilo, Phys. Rev. Lett. {\bf 103}, 210401 (2009).

%
\bibitem{Haikka} P. Haikka, J. Goold, S. McEndoo, F. Plastina, and S. Maniscalco, Phys. Rev. A {\bf 85}, 060101(R) (2012).
%

\bibitem{theoryJie} J. Li, {\it et al.}, {\it On the link between dynamical non-Markovianity  and extended Bell's inequalities in time}, (to appear, 2013).


\bibitem{SusanaSantos} S. F. Huelga, T. W. Marshall, and E. Santos, Phys. Rev. A {\bf 52}, R2497 (1995).

\bibitem{SusanaSantos2} S. F. Huelga, T.W. Marshall, and E. Santos, Phys. Rev. A {\bf 54}, 1798 (1996).

\bibitem{bohm} J. S. Bell, Physics {\bf 1}, 195 (1964). 

\bibitem{fidelityNMR} X. Wang, C.-S. Yu, and X. X. Yi, Phys. Lett. A {\bf 373}, 58 (2008).
\bibitem{tomoNMR} J. Maziero, R. Auccaise, L. C. Celeri, D. O. Soares-Pinto, E. R. deAzevedo, T. J. Bonagamba, R. S. Sarthour, I. S. Oliveira, and R. M. Serra, Braz. J. Phys. {\bf 43}, 86 (2013).
\bibitem{kofler} J. Kofler, and C. Brukner, Phys. Rev. Lett. {\bf 99}, 180403 (2007). 
\bibitem{leggett}  A. J. Leggett and A. Garg, Phys. Rev. Lett. {\bf 54}, 857 (1985).
\bibitem{refDiogo} A. Abragam, {\it The Principles of Nuclear Magnetism} (Oxford University, New York, 1978); R. R. Ernst, G. Bodenhausen, and A. Wokaum, {\it Principles of Nuclear Magnetic Resonance in One and Two Dimensions} (Oxford University, New York, 1987).

\end{thebibliography}
\end{document}